\documentclass[sigconf,balance=false]{acmart}
\usepackage{popets}

\usepackage{microtype}
\usepackage{multirow}
\usepackage{subcaption} 

\setcopyright{popets}
\copyrightyear{YYYY}

\acmYear{YYYY}
\acmVolume{YYYY}
\acmNumber{X}
\acmDOI{XXXXXXX.XXXXXXX}
\acmISBN{}
\acmConference{Proceedings on Privacy Enhancing Technologies}
\begin{document}

\title[Revisiting the LiRA Membership Inference Attack Under Realistic Assumptions]{Revisiting the LiRA Membership Inference Attack Under Realistic Assumptions}



\author{Najeeb Jebreel}
\orcid{0000-0002-4911-3802}
\affiliation{%
  \institution{Universitat Rovira i Virgili, Catalonia}
  \city{}
  \state{}
  \country{}}
\email{najeeb.jebreel@urv.cat}

\author{Mona Khalil}
\affiliation{%
  \institution{Universitat Rovira i Virgili, Catalonia}
  \city{}
  \state{}
  \country{}}
\email{mona.khalil@urv.cat}

\author{David Sánchez}
\affiliation{%
  \institution{Universitat Rovira i Virgili, Catalonia}
  \city{}
  \state{}
  \country{}}
\email{david.sanchez@urv.cat}

\author{Josep Domingo-Ferrer}
\affiliation{%
 \institution{Universitat Rovira i Virgili, Catalonia}
 \city{}
  \state{}
  \country{}}
\email{josep.domingo@urv.cat}


\renewcommand{\shortauthors}{Jebreel et al.}

\begin{abstract}
    Membership inference attacks (MIAs) have become the standard tool for evaluating privacy leakage in machine learning (ML).
    Among them, the \emph{Likelihood-Ratio Attack} (LiRA) is widely regarded as the state of the art when sufficient shadow models are available.
    However, prior evaluations have often overstated
    the effectiveness of LiRA by attacking models overconfident on their training samples, calibrating thresholds on target data, assuming balanced membership priors, and/or overlooking attack reproducibility.
    We re-evaluate LiRA under a \emph{realistic} protocol that
    (i) trains models using anti-overfitting (AOF) (and transfer learning (TL), when applicable) to reduce overconfidence as it would be desirable in production models;
    (ii) calibrates decision thresholds from shadow models and data rather than (usually unavailable) target data;
    (iii) measures positive predictive value (PPV, a.k.a. precision) under shadow-based thresholds and skewed---rather than unrealistically balanced---membership priors ($\pi \le 10\%$); and
    (iv) quantifies per-sample membership reproducibility across different seeds and training variations.
    In this setting, we find that
    (a) AOF significantly weakens LiRA and TL further reduces the effectiveness of the attack, while improving model accuracy;
    (b) with shadow-based thresholds and skewed priors, 
    LiRA’s PPV often drops from near-perfect to substantially lower levels, especially under AOF/AOF+TL and for $\pi \le 10\%$; and
    (c) LiRA’s thresholded ``vulnerable'' sets at extremely low FPR exhibit poor reproducibility across runs, while likelihood ratio-based rankings are more stable.
    These results suggest that
    (i) LiRA, and likely weaker MIAs, are less effective than previously suggested, and their positive inferences can be less reliable under realistic settings; and
    (ii) for MIAs to serve as meaningful privacy auditing tools, their evaluation must reflect pragmatic training practices, feasible attacker assumptions, and reproducibility considerations.
    We release our code at: \url{https://github.com/najeebjebreel/lira_analysis}.
\end{abstract}

\keywords{machine learning, privacy, membership inference attacks, attack reliability, reproducibility}

\maketitle

\section{Introduction}
\label{sec:introduction}

Machine learning (ML) models are frequently trained or fine-tuned on personal data from sensitive domains such as healthcare~\cite{kaissis2020secure}, finance~\cite{gogas2021machine}, and law~\cite{cui2023chatlaw}.
This raises privacy concerns for both individuals and regulators,
 as trained models can leak information about their training data~\cite{papernot2018sok,rigaki2023survey}.
Membership inference attacks (MIAs)~\cite{homer2008resolving,shokri2017membership} embody one of these concerns: By inferring whether a specific sample was part of a model training set, an attacker could infer sensitive information (\emph{e.g.}, the participation of an individual
in a medical study can indicate that
the individual suffers from the disease under study).
In this regard, some standardization bodies now classify MIAs as a threat to training data confidentiality~\cite{tabassi2019taxonomy,murakonda2020ml}.
Moreover, due to their simplicity and agnostic nature, MIAs are widely used to empirically assess training data leakage in ML and unlearning~\cite{shokri2017membership,melis2019exploiting,lukas2023analyzing,kurmanji2023towards,blanco2025digital}.

\paragraph{Why LiRA} The Likelihood-Ratio Attack (LiRA)~\cite{carlini2022membership} is widely regarded as the state of the art in MIAs.
Multiple recent studies have shown that LiRA consistently dominates alternatives in the extremely low false positive rates (FPRs) regime~\cite{li2024privacy,shi2024learning,hayes2025strong,pollock2025free,bai2025empirical}, where reliable inference matters~\cite{aerni2024evaluations}.
Although newer methods (\emph{e.g.}, Attack R~\cite{ye2022enhanced}, RMIA~\cite{zarifzadeh2024lowcost}, or LDC-MIA~\cite{shi2024learning}) can achieve a higher true positive rate (TPR) with limited resources, they do not outperform LiRA when sufficient shadow models are available to the attacker (\emph{i.e.}, $> 128$).
Specifically, \cite{pollock2025free} shows that at an FPR of 0.1\%, LiRA identifies substantially more vulnerable samples than RMIA~\cite{zarifzadeh2024lowcost} or Attack R~\cite{ye2022enhanced}, and detects nearly all 
samples labeled as vulnerable by its competitors; 
\cite{li2024privacy} show that LiRA is effective in capturing memorization-based leakage beyond simple overfitting.

We focus on LiRA as a privacy benchmark for two reasons:
(i)~weaker attacks are likely to perform similarly or worse than LiRA;
(ii)~prior evaluations of LiRA have adopted optimistic choices that may have overestimated its effectiveness, including:
\begin{enumerate}
\item \textbf{Loss-wise overfitting.} Target models typically exhibit a large test-to-train accuracy gap~\cite{carlini2022membership,liu2022membership,zarifzadeh2024lowcost,wang2025membership}, or a large corresponding \emph{loss ratio} even when the accuracy gap seems to be small (see~\autoref{tab:utility}), reflecting overconfidence in training data that facilitate MIAs~\cite{overconfidence2024,zhu2025impact}.
These gaps can be reduced with anti-overfitting (AOF) techniques~\cite{dionysiou2023sok,blanco2022critical}, which are standard in production settings as they improve model generalization.
\item \textbf{Target-based thresholds.} Membership \emph{thresholds} are derived on the (usually unavailable) target model's labeled data. This overfits the decision threshold and unrealistically benefits external attackers.
\item \textbf{Balanced priors.} Evaluations assume a balanced membership prior ($\pi = 50\%$), while members, in most cases, are a small set of a larger population~\cite{jayaraman2021revisiting,all2024genomic}.
\item \textbf{Overlooked reproducibility.} Per-sample reproducibility across different seeds or training variations is often overlooked, leaving open the question of whether attack inferences are stable and reliable.
\end{enumerate}

Moreover, evaluations typically disregard the increasing use of \emph{transfer learning (TL)} in privacy-sensitive domains with limited data~\cite{kim2022transfer,yin2019feature,lanzetta2024transfer}, although TL generally improves both model utility and robustness to MIAs compared to training from scratch~\cite{he2021quantifying,rajabi2024poster,bai2025empirical}.

This background motivates our four practical questions:
\textbf{(Q1)} How do AOF and TL affect the utility and effectiveness of LiRA?
\textbf{(Q2)} What is the effect of shadow-only threshold calibration on attack 
effectiveness and reliability?
\textbf{(Q3)} How do the skewed membership priors (\emph{e.g.}, $\pi \le 10\%$) change PPV at low FPR?
\textbf{(Q4)} How reproducible are LiRA outputs across training runs?

\paragraph{Contributions}
To answer these questions, we reassess LiRA under realistic ML practice and a strong but realistically constrained black-box attacker, with reproducibility in mind.
Specifically, 
we assume a well-resourced attacker capable of training multiple shadow models on data from the same distribution as the target’s training set~\cite{shokri2017membership,carlini2022membership}.
We consider the model owner as a pragmatic ML practitioner, who employs anti-overfitting and, when applicable, transfer learning to improve model utility.
This setting captures a conservative upper bound on the black-box membership leakage that can be achieved by a strong attacker against well-trained models.

Specifically, our contributions are as follows.
\begin{itemize}
\item We design a comprehensive evaluation protocol that (i) systematically varies defender practices (data augmentation, regularization, transfer learning) and attacker assumptions (threshold calibration, membership priors), and (ii) defines consistent evaluation metrics covering \emph{attack effectiveness} (TPR@low-FPR), \emph{reliability} (PPV under realistic priors), 
and \emph{reproducibility across runs} at both levels: (i) thresholded sets inferred at a target FPR and (ii) ranked samples based on granular vulnerability scores (likelihood ratios).

\item We show that combining AOF techniques significantly weakens LiRA while preserving model utility, and that TL offers further protection and enhances model utility.

\item We show that under shadow-based threshold calibration and realistically skewed priors ($\pi \leq 10\%$), the achieved FPRs deviate from nominal targets, often causing LiRA’s PPV to drop from near-perfect (under target-calibrated thresholds) to substantially lower levels. This makes positive inferences less reliable, especially for $\pi \le 10\%$.

\item We quantify reproducibility across training randomness, showing that LiRA’s thresholded ``vulnerable'' sets at extremely low FPR are highly unstable; in contrast, likelihood ratio-based rankings are more stable than thresholded sets, although single-run top-$q\%$ identification remains unstable for small $q$.

\item We identify a strong relationship between the test-to-train loss ratio and LiRA's success, which provides a lightweight, attack-free proxy for monitoring empirical privacy risk. This ratio reflects the global difference in prediction confidence (and, indirectly, uncertainty) between member and non-member data.
\end{itemize}

Our findings indicate that LiRA (and likely weaker MIAs) are significantly less effective, reliable, and reproducible under realistic conditions than previously suggested.
Our evaluation protocol provides concrete guidance for realistic and meaningful privacy audits, emphasizing that MIA evaluations must reflect pragmatic training practices, feasible attacker assumptions, and reproducibility considerations.

\section{Background}
\label{sec:background_related}

\subsection{Membership Inference Attacks and LiRA}
\label{sec:attacks}

Membership inference attacks (MIAs) aim to determine whether a sample $(x,y)$ was present in the training set $D_{\text{train}}$ of a target model $f_{\theta}$, 
where $x \in \mathcal{X}$ and $y \in \mathcal{Y}$, with $\mathcal{X}$ being the input space and $\mathcal{Y}$ the label space.  
Since the pioneering work of~\cite{shokri2017membership}, which introduced the MIA based on ``shadow models'', several black-box MIAs have been proposed.
These methods exploit different signals derived from the model output when queried with a sample, ranging from raw loss or prediction confidence to calibrated, sample-specific scores, or even the final prediction label~\cite{yeom2018privacy,salem2019ml,ye2022enhanced,watson2022on,carlini2022membership,bertran2023scalable,zarifzadeh2024lowcost,choquette2021label}.

White-box MIAs, which exploit access to model internals, have also been proposed~\cite{nasr2019comprehensive,leemann2024my} and can sometimes yield marginal gains over black-box methods~\cite{leemann2024my}.  
However, black-box attacks remain the standard benchmark for assessing membership and are considered more relevant because (i) it is more feasible that attackers can observe only model outputs, and (ii) they are nearly as effective as white-box variants, as the final loss of samples has been shown to be the strongest single predictor of membership~\cite{sablayrolles2019white,leemann2024my}.

LiRA~\cite{carlini2022membership} is widely regarded as the state-of-the-art black-box MIA at extremely low FPRs~\cite{shi2024learning,hayes2025strong,pollock2025free,bai2025empirical}. 
It formulates membership inference as a \emph{statistical hypothesis test} that accounts for \emph{per-sample difficulty}~\cite{sablayrolles2019white}: for a candidate sample $(x,y)$ and target model $f_{\theta}$, the attacker tests whether $(x,y) \in D_{\text{train}}$ versus $(x,y) \notin D_{\text{train}}$ using shadow models to approximate the distributions of outputs for members versus non-members.
Shadow models trained with and without $(x,y)$ produce score distributions that LiRA transforms and models as $\mathcal{N}(\mu_{\text{IN}}, \sigma_{\text{IN}}^2)$ and $\mathcal{N}(\mu_{\text{OUT}}, \sigma_{\text{OUT}}^2)$. 
LiRA transforms the sample's true-class confidence $p = f_{\theta}(x)_y$ using $\phi(p) = \log(p/(1-p))$, which yields approximately Gaussian distributions.

\emph{Online LiRA} computes the likelihood ratio between the IN and OUT distributions for the observed score $\phi_{\text{obs}} = \phi(f_{\theta}(x)_y)$, and declares membership when this ratio exceeds a threshold $\tau$ chosen to control FPR.
Beyond thresholded decisions, LiRA’s per-sample likelihood ratios provide a continuous vulnerability signal that can support ranking-based analyses across runs.
\emph{Offline LiRA} reduces computational cost by using only OUT shadow models, evaluating the right-tail probability $p_{\text{out}} = \Pr[Z \geq \phi_{\text{obs}} \mid Z \sim \mathcal{N}(\mu_{\text{OUT}}, \sigma_{\text{OUT}}^2)]$ and declaring membership when $p_{\text{out}}$ falls below the target FPR. 
Multivariate distributions using multiple augmentations of the input (\emph{e.g.}, shifts, flips) improve attack effectiveness.
Variance estimation can be per sample or fixed globally. 
With many shadow models ($\geq 64$), the authors find that the per-sample variances yield stronger results, and the fixed variance (FV) is more stable with fewer models. 

\subsection{Defenses against MIAs}
\label{sec:defenses}

Defenses against MIAs fall into two broad categories: \emph{formal} methods with certified privacy guarantees, such as differential privacy (DP), 
and \emph{empirical} approaches, which mitigate leakage in practice but lack formal guarantees, such as standard regularization and data augmentation~\cite{truex2021demystifying,hu2022membership}.

\paragraph{Differential privacy (DP~\cite{dwork2006differential})} DP in ML is typically implemented via DP-SGD~\cite{abadi2016deep}, 
which clips per-sample gradients and injects Gaussian noise during training to bound the influence of any single record. 
With a meaningful privacy budget (\emph{i.e.}, $\varepsilon \leq 1 $ \cite{dwork2011firm}), DP provides formal guarantees by making membership inference statistically unreliable. 
However, beyond its substantial computational overhead, DP inevitably degrades the utility of models~\cite{blanco2022critical,wang2023differential,du2025can}, restricting its adoption in real-world deployments~\cite{practice}.

\paragraph{Empirical defenses}
Early work has established a connection between overfitting and MIA vulnerability~\cite{shokri2017membership,yeom2018privacy}. 
This motivates the use of standard anti-overfitting techniques (such as early stopping~\cite{caruana2000overfitting}, weight decay~\cite{krogh1991simple}, dropout~\cite{srivastava2014dropout}, and data augmentation~\cite{cubuk2018autoaugment}) which have been shown to mitigate leakage by reducing generalization gaps~\cite{shokri2017membership,salem2019ml,sablayrolles2019white,kaya2021does,blanco2022critical}. 
Empirical studies~\cite{kaya2021does,blanco2022critical,lomurno2022utility} demonstrate that these techniques (especially when combined) can substantially reduce MIA success rates.
Another line of defense limits the information revealed at inference time~\cite{shokri2017membership,jia2019memguard}. 

A related technique is transfer learning~\cite{pan2009survey}, which adapts models pretrained on large-scale datasets (\emph{e.g.}, ImageNet, large text corpora) to smaller, task-specific data via fine-tuning.  
Unlike training from scratch —where limited datasets often cause overfitting and increase susceptibility to MIAs—, TL reuses broad, stable features and requires fewer samples and epochs, reducing dataset-specific memorization.  
TL is now widely adopted in domains where accuracy is critical but data are scarce, such as healthcare~\cite{kim2022transfer,ebbehoj2022transfer}, 
 making it relevant for realistic MIA evaluation.  
Empirical work also confirms its protective effect: fine-tuned models are less vulnerable than scratch-trained ones~\cite{he2021quantifying,rajabi2024poster,bai2025empirical}. 
Specifically, \cite{he2021quantifying} show that pretraining reshapes the privacy–utility landscape, 
\cite{rajabi2024poster} find that combining TL with randomization weakens label-only MIAs, 
and~\cite{bai2025empirical} provide evidence of reduced attack success under TL.

\section{Related Work}
\label{sec:related}

A growing body of work has raised concerns about the evaluation of MIAs, including LiRA. 

\paragraph{Dependence on overfitting and poor calibration}
The success of MIAs mainly depends on overfitting and miscalibration, which cause models to be overconfident in their training samples. 
This facilitates the separation between members and non-members~\cite{dionysiou2023sok,overconfidence2024,zhu2025impact}.
\cite{zhu2025impact} show theoretically that the advantage of LiRA (and LiRA-style attacks) grows with \emph{model miscalibration}---when predicted confidence mismatches the true accuracy---and decreases with both data and model uncertainty. 
\cite{dionysiou2023sok} demonstrate that when models are trained to achieve high test accuracy and small generalization gaps, all major MIAs, including LiRA, lose most of their effectiveness.
AOF techniques have been shown to substantially reduce MIA success with a better privacy-utility trade-off than DP~\cite{kaya2021does,blanco2022critical,lomurno2022utility}. \cite{zhang2024membership} show that modified MixUp augmentation for vision transformers substantially reduces attention-based membership leakage while preserving or improving model utility.
In TL settings, score-based MIA success (including LiRA) declines sharply as model generalization improves~\cite{bai2025empirical}, 
and even fine-tuned LLMs such as BERT experience reduced attack advantage, especially in terms of TPR at low FPR~\cite{he2024difficulty}.

\paragraph{Reliability and reproducibility}
\cite{rezaei2021difficulty} show that naturally trained deep learning (DL) models tend to behave similarly on training and non-training samples, causing MIAs to produce high FPRs. \cite{hintersdorf2022trust} demonstrate that the overconfidence of modern DL models leads score-based MIAs to produce many false positives.
Due to several training and attack randomization factors, different attacks (or even different instances of the same attack) often produce highly non-overlapping subsets of ``vulnerable'' (\emph{i.e.}, member) samples despite similar global metrics~\cite{wang2025membership,zhang2025position,ye2022enhanced}. 
\cite{wang2025membership} report that six state-of-the-art MIAs, including LiRA, exhibit low consistency at the sample level between runs (Jaccard $<0.4$ at $0.1\%$ FPR), with only the deterministic loss-based attack~\cite{yeom2018privacy} maintaining consistency.
Similar behavior was also observed by~\cite{ye2022enhanced}.
This non-reproducibility limits the use of a single run to confidently identify a small, stable subset of ``vulnerable'' records for targeted mitigation~\cite{jiacheng2024mist}.

\paragraph{Evaluation assumptions}
Assuming balanced membership priors (50/50) can highly inflate the perceived privacy risk, as training members usually represent a small fraction of the overall population (especially in privacy-sensitive domains). 
Even a modest FPR can generate many false positives and degrade PPV under skewed priors~\cite{rezaei2021difficulty,jayaraman2021revisiting}. 
\cite{jayaraman2021revisiting} emphasize the use of PPV under realistically skewed priors for a realistic evaluation of MIA.
Moreover, most evaluations overestimate attack success by tuning decision thresholds directly on the scores computed from the target model on its labeled data, giving attackers an unrealistic advantage in choosing the optimal membership threshold. 
\cite{he2024difficulty} note that shadow models can be used to choose thresholds to achieve a target FPR, but show that such shadow-derived thresholds transfer poorly across datasets and model architectures. 
\cite{ye2022enhanced} emphasize that, to measure genuine leakage rather than artifacts of prior mismatch, the data used to evaluate MIAs should come from a distribution similar to that of the target model’s training data.
Otherwise, results may over- or under-estimate the true leakage.
Finally, LiRA's evaluation protocol assumes a powerful attacker with near-perfect auxiliary data and the ability to replicate the target's architecture and training pipeline~\cite{shokri2017membership,liu2022membership,salem2019ml}.
While defining worst-case bounds, this overstates realistic leakage: even mild shadow-target mismatches can affect score distributions and inflate FPR~\cite{he2024difficulty,fu2024membership,zhu2025impact}.

\section{Evaluation Protocol and Experimental Setup}
\label{sec:protocol_setup}

Although previous work has highlighted MIA (and LiRA) limitations, it has addressed them in isolation. 
Such isolated evaluations can miss compounded effects that become apparent only under joint conditions.
We evaluate LiRA under \emph{realistic joint} conditions by
(i) training less overfitted models with lower prediction-confidence certainty while maintaining accuracy; 
(ii) calibrating decision thresholds using only shadow models and data; 
(iii) evaluating PPV under shadow-based thresholds and realistic membership priors ($\pi \le 10\%$); 
and (iv) quantifying \emph{per-sample} variability across runs and training randomness, both for thresholded membership sets and for score/rank stability.
{\em This integrated evaluation clarifies how methodological choices (training practices, threshold calibration, priors, run-to-run variability) affect measured privacy leakage.}

\subsection{Threat Model}

\paragraph{Attacker} 
Following prior work~\cite{shokri2017membership,carlini2022membership,zarifzadeh2024lowcost}, we assume that the attacker has \emph{black-box} query access to a deployed target model $f_\theta$ and aims to infer whether a target sample $(x,y)$ was included in $f_\theta$'s training set.
Although LiRA entails a high computational cost (since its effective performance requires training and querying hundreds of shadow models), we do not restrict the resources allocated to the attack. 
We assume a well-resourced attacker capable of training 256 \emph{shadow models}, as considered in the original LiRA paper~\cite{carlini2022membership}. 

The attacker can train shadow models on data drawn from the same distribution as the target's training data, and approximate the target's architecture and hyperparameters~\cite{tramer2016stealing,bastani2017interpreting,wang2018stealing}.
Notice that the models obtained will inevitably differ due to randomness in the data distribution, stochastic training, or potential small variations in hyperparameters.  
Importantly, we enforce the following two realistic constraints. 
(1) The attacker cannot calibrate the decision threshold from the scores obtained from the target model on its training data. This access would trivialize the attack by directly revealing membership of all training data, which is inconsistent with a realistic black-box threat model. 
(2) The attacker cannot assume balanced membership priors, as real training data are, most of the time, a small fraction of a larger population, especially in sensitive domains~\cite{murthy2004participation,all2024genomic}.

Instead, thresholds are calibrated exclusively on shadow models, with posterior beliefs adjusted via the Bayes' rule using realistic priors $\pi$~\cite{jayaraman2021revisiting,song2021systematic}.
When $\pi$ is unknown, plausible priors should be considered.
This constitutes a \emph{strong yet realistic} attacker: strong enough to approximate an upper bound on black-box risk while respecting realistic constraints.

\paragraph{Defender} 
The defender is modeled as a pragmatic ML practitioner who aims to deploy accurate models while simultaneously minimizing privacy leakage.  
Since MIAs primarily exploit overfitting~\cite{yeom2018privacy}, we assume the defender applies standard AOF techniques (data augmentation, weight decay, dropout and/or early stopping), and TL when applicable.
These techniques offer not only favorable privacy--utility trade-offs~\cite{blanco2022critical}, but also improve the generalizability of the model, which is desirable in production settings.
Thus, we assume that the defender has an incentive to rely on tuned combinations of these techniques, which can be achieved through automated hyperparameter optimization~\cite{snoek2012practical,falkner2018bohb}.

\subsection{Datasets and Models}
\label{sec:data_models}
We evaluated LiRA on four datasets. 
{\em CIFAR-10/100}~\cite{krizhevsky2009learning} each contain 60,000 images of size \(32\times32\) with 10/100 classes. 
{\em GTSRB}~\cite{stallkamp2011german} consists of 51,839 traffic sign images (43 classes). 
{\em Purchase-100}~\cite{shokri2017membership} includes 197,324 purchase records, each with 600 features and 100 classes.

The CIFAR-10/100 and Purchase-100 datasets were used in the LiRA paper~\cite{carlini2022membership}, while GTSRB was examined in~\cite{liu2022membership}. 
We focus on the CIFAR datasets because, as shown in~\cite{carlini2022membership}, CIFAR-10 exhibited vulnerability despite a small train-test accuracy gap, and CIFAR-100 showed the highest leakage among the image datasets. 
On the other hand, the Purchase-100 dataset is widely adopted in MIA research, but~\cite{carlini2022membership} argue that its simplicity makes it a poor proxy for realistic privacy risk assessment; we include it here for completeness. 
GTSRB serves as a realistic benchmark as it achieves near-perfect accuracy with a minimal train--test loss ratio, making it representative of well-calibrated, high-utility models likely to be deployed in practice.

All input features were normalized using dataset-specific statistics. 
We used ResNet-18~\cite{he2016deep} for CIFAR-10 and GTSRB, WideResNet~\cite{zagoruyko2016wide} for CIFAR-100, and a fully connected network (FCN)~\cite{carlini2022membership} for Purchase-100, all trained from scratch. 
Furthermore, we used EfficientNet-V2~\cite{tan2021efficientnetv2} for transfer learning experiments.

\subsection{Training Configurations}
\label{sec:training_configs}
We define three benchmarks with increasing regularization strength:

\paragraph{Baseline} This benchmark replicates the original setup of LiRA~\cite{carlini2022membership}: image models trained using three image augmentations (horizontal flip, random crop with reflection padding, Cutout~\cite{devries2017improved}), no dropout, and weight decay $5{\times}10^{-4}$.
They were trained from scratch with SGD with momentum 0.9, cosine scheduling with an initial learning rate 0.1~\cite{loshchilov2017sgdr}, and a batch size 256.
The FCN models used the same settings, but with an initial learning rate of 0.01 and a batch size of 128 without data augmentation.
All models were trained from scratch for up to 100 epochs with early stopping after 20 epochs of no improvement.
Utility and LiRA evaluations were performed using the best saved checkpoints based on validation accuracy.

\paragraph{Anti-overfitting}
These benchmarks combine comprehensive AOF: multiple image augmentations (horizontal flip, random crop with reflection padding, rotation, ColorJitter, CutMix~\cite{yun2019cutmix}), dropout (10\% vision, 50\% tabular) and/or increased weight decay ($10^{-3}$).
These configurations achieved favorable utility--privacy trade-offs with test-to-train loss ratios below 2.0 while maintaining accuracy close to that of baseline.
Table~\ref{tab:configs} summarizes the configurations; 
see Appendix~\ref{app:ablation} for a complete ablation study and additional details on training configurations.
\newcommand{\cmark}{\checkmark}
\begin{table}[t!]
\centering\scriptsize
\caption{Benchmark configurations. FLP=H.Flip, CRP=R.Crop with padding, ROT=Rotation, JTR=ColorJitter, CUT=Cutout, CMX=CutMix, DRP=dropout ratio, WD=weight decay.}
\label{tab:configs}
\resizebox{\linewidth}{!}{
\begin{tabular}{@{}llccccccccc@{}}
\toprule
Dataset & Benchmark & Architecture & FLP & CRP & ROT & JTR & CUT & CMX & DRP (\%) & WD \\
\midrule
\multirow{3}{*}{CIFAR-10}
  & baseline & ResNet-18       & \cmark & \cmark &      &      & \cmark &      & 0  & 5e-4 \\
  & AOF     & ResNet-18       & \cmark & \cmark & \cmark & \cmark &      & \cmark & 10 & 1e-3 \\
  & TL       & EfficientNet-V2 & \cmark & \cmark & \cmark & \cmark &      & \cmark & 25 & 5e-2 \\
\cmidrule{1-11}
\multirow{3}{*}{CIFAR-100}
  & baseline & WideResNet      & \cmark & \cmark &      &      & \cmark &      & 0  & 5e-4 \\
  & AOF     & WideResNet      & \cmark & \cmark & \cmark & \cmark &      & \cmark & 10 & 1e-3 \\
  & TL       & EfficientNet-V2 & \cmark & \cmark & \cmark & \cmark &      & \cmark & 25 & 5e-2 \\
\cmidrule{1-11}
\multirow{2}{*}{GTSRB}
  & baseline & ResNet-18       & \cmark & \cmark &      &      &      &      & 0  & 5e-4 \\
  & TL       & EfficientNet-V2 & \cmark & \cmark & \cmark & \cmark &      & \cmark & 25 & 5e-2 \\
\cmidrule{1-11}
\multirow{2}{*}{Purchase-100}
  & baseline & FCN             &        &        &      &      &      &      & 0  & 5e-4 \\
  & AOF     & FCN             &        &        &      &      &      &      & 50 & 1e-3 \\
\bottomrule
\end{tabular}}
\end{table}

\paragraph{AOF + TL}
We fine-tuned ResNet-18 and EfficientNet-V2-S~\cite{tan2021efficientnetv2} pretrained on ImageNet~\cite{deng2009imagenet} for 10 and 5 epochs using AdamW with OneCycle scheduling~\cite{smith2019super}, batch size 64--128, dropout 25\%, weight decay $5{\times}10^{-2}$, and the data augmentations used above.
EfficientNet-V2-S offered higher utility while delivering comparable resistance to LiRA (see Table~\ref{tab:cifar10-tl_effect} in Appendix~\ref{app:tl_choice}); thus, we adopted EfficientNet-V2-S for our TL configurations.

\smallskip
\noindent Additional training details are presented in Appendix~\ref{app:additional_setting}.

\subsection{Attack Configuration}
\label{sec:attack_instantiation}

Following the original LiRA setting~\cite{carlini2022membership}, we trained the $M{=}256$ shadow models for each benchmark to obtain the IN/OUT score distributions.  
We constructed balanced member/non-member splits so each model (shadow/target) contains roughly half of the dataset as members, and every sample is a member in exactly 128 shadow models.  
In inference, following~\cite{carlini2022membership}, we adopted 18 deterministic transformed inputs for images (original, horizontal flip, and eight 2-pixel shifts with their reflections).
For tabular data, a single query per sample was used.  
We evaluated both \emph{online} (IN/OUT) and \emph{offline} (OUT-only) LiRA variants, each with per-sample and fixed variance (FV) estimation.  
We also considered the global thresholding attack, which computes threshold scores of the target model without shadow training to show how average-case metrics (\emph{e.g.}, Acc/AUC) can be misleading.

\subsection{Threshold Calibration}

\paragraph{Optimistic}
A per-target threshold~$\tau(\alpha)$ is computed directly from the predictions of the target model on its own data to achieve an optimal nominal FPR~$\alpha$.  
This setting, used by~\cite{carlini2022membership} and most existing MIAs, assumes privileged access and therefore represents an upper bound on attack performance from the auditor’s perspective.

\begin{figure}[t!]
  \centering
  \begin{subfigure}[b]{0.5\linewidth}
    \centering
    \includegraphics[width=\linewidth]{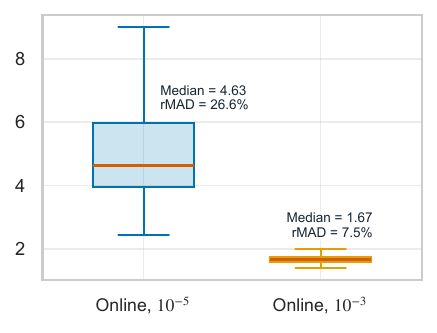}
    \caption{Single run}
    \label{fig:thresh_calibration_single}
  \end{subfigure}
  \begin{subfigure}[b]{0.48\linewidth}
    \centering
    \includegraphics[width=\linewidth]{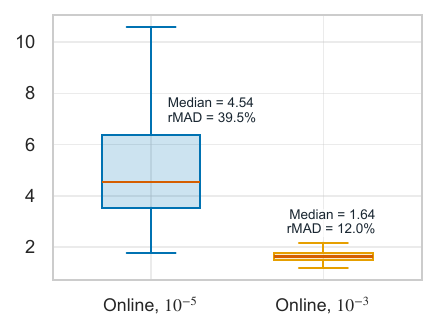}
    \caption{12 independent runs}
    \label{fig:thresh_calibration_multi}
  \end{subfigure}
  \caption{Distributions of decision thresholds of the \emph{online} LiRA variant at nominal FPRs of $0.001\%\,(10^{-5})$ and $0.1\%\,(10^{-3})$ on CIFAR-10 (ResNet-18, AOF).
        Boxes show interquartile ranges with medians and relative median absolute deviations (rMAD) from 256 shadow models of a single run (a) and 12 independent runs (b).
        Threshold variability is substantially higher at $0.001\%$ than at $0.1\%$, indicating unstable calibration at the extreme tail required for high inference precision.}
        \Description{Two side-by-side boxplots showing decision threshold distributions
        for the online LiRA variant at nominal FPRs of $10^{-5}$ ($0.001\%$) and $10^{-3}$ ($0.1\%$).
        Left: single-run results. Right: aggregated across 12 independent runs.
        Variability increases as the target FPR decreases, indicating reduced threshold stability at the extreme tail.}
  \label{fig:thresh_calibration}
\end{figure}

\paragraph{Realistic}
Here, thresholds are estimated exclusively from the scores obtained from the predictions of shadow models on their respective shadow data.
As described below, we use each of the $M$ shadow
models as the target once, and estimate its membership threshold as the \emph{median} over 
the remaining $M-1$ shadows, that is,
$\tau_{\text{shadow}}(\alpha) = \mathrm{median}\bigl(\{\tau_i(\alpha)\}_{i \neq \text{target}}\bigr)$,
where $\tau_i(\alpha)$ is the target FPR $\alpha$ on shadow~$i$.
The median provides a robust estimate while limiting the influence of outliers.
Fig.~\ref{fig:thresh_calibration} shows a larger threshold variability between targets of the same run, which increased further with additional runs.
This variability is expected to alter both the achieved TPR and FPR compared to the controlled optimistic setting.
It also highlights that pursuing high precision at ultra-low FPR introduces costly instability, whereas relaxing the FPR yields more reproducible thresholds, and this is  expected to sacrifice inference precision under realistic skewed membership priors.

\subsection{Evaluation Metrics}
\label{sec:evaluation_metrics}
\paragraph{Evaluation mode}
\cite{carlini2022membership} use one shadow model from each run as the target.  
In contrast, we adopt a \emph{leave-one-out} mode, where each of the $M$ shadows serves as a target once, with results aggregated across all targets.  
This captures variability among potential target models and eliminates the selection bias that single-target evaluation obscures.

\paragraph{Utility metrics}
We measure training and test accuracies and the mean cross-entropy loss, and report their averages across target models.
We define the \emph{loss ratio} as $ \mathrm{LR}=L_{\text{test}}/L_{\text{train}},$
which captures confidence-based generalization gaps that MIAs exploit more directly than the 0/1 accuracy gaps~\cite{li2021membership}.

\paragraph{Privacy metrics}
Following~\cite{carlini2022membership,zarifzadeh2024lowcost}, we report TPR at extremely low FPRs ($\alpha \in \{0.001\%, 0.1\%\}$) where the inferences are more trustworthy.
For a realistic assessment, we compute the positive predictive value (PPV) under skewed priors~\cite{jayaraman2021revisiting}:
$\text{PPV}(\pi) = \pi \cdot \text{TPR'}/(\pi \cdot \text{TPR'} + (1-\pi) \cdot \text{FPR}'),$
where FPR$'$ is the achieved (not nominal) FPR, and TPR$'$ is the corresponding achieved TPR.
PPV is relevant as it indicates how certain an attacker can be about positive inferences under realistic priors. 
For these privacy metrics, we report the mean~$\pm$~standard deviation across target models.

\paragraph{Reproducibility metrics}
Following~\cite{wang2025membership}, we assess per-sample reproducibility across $K$ independent runs by comparing the sets of samples inferred as members at a threshold corresponding to a target FPR.
Across runs, the training/test split is fixed; variability arises from shadow partitioning, random initialization, and stochastic training dynamics (\emph{e.g.}, batch shuffling).
For each pair of runs, we compute the Jaccard similarity $J(A,B)=|A\cap B|/|A\cup B|$, averaged over all $\binom{K}{2}$ pairs, and also report mean $|A\cap B|$ and $|A\cup B|$.
To go beyond pairwise comparisons, we additionally consider $k$-wise agreement over each subset of $k$ runs and average these quantities over all $\binom{K}{k}$ subsets, including $k=K$.
In addition, we focus on the most vulnerable samples: for each run $r$, we define $S_r$ as the subset of samples flagged as members with \emph{zero per-sample false positives} (0FP) and within-run support \emph{TP$\geq x$} (detected by at least $x$ of the 128 IN shadows).
We further evaluate score-level stability. For each run $r$ and sample $x$, we define a vulnerability score using the IN/OUT shadow-model log-likelihood ratios $\Lambda_{\text{IN}}^{(r)}(x)$ and $\Lambda_{\text{OUT}}^{(r)}(x)$ via the \texttt{median gap}
$\Delta_{\text{med}}^{(r)}(x)=\mathrm{median}[\log \Lambda_{\text{IN}}^{(r)}(x)]-\mathrm{median}[\log \Lambda_{\text{OUT}}^{(r)}(x)]$.
Samples are ranked by $\Delta_{\text{med}}^{(r)}(\cdot)$, and $T_q^{(r)}$ denotes the top-$q\%$ set in run $r$.
For each run pair $(r,r')$, we report set agreement ($|{\cap}|$, $|{\cup}|$, Jaccard) and tail rank stability (Spearman on $T_q^{(r)} \cap T_q^{(r')}$), averaged over all $\binom{K}{2}$ pairs; we also report the corresponding all-runs ($k=K$) metrics.

\section{Results}
\label{sec:results}

\subsection{Impact of AOF and TL}
\label{sec:aof_tl}

\paragraph{Model utility} 
Table~\ref{tab:utility} presents model utility metrics across benchmarks. 
AOF techniques consistently achieved an accuracy comparable to that of baseline models. 
TL delivered the highest accuracy across all datasets where applied. 
With CIFAR-100, for example, TL provided an improvement of $+14.26\%$ in the test accuracy.
We can see that the baseline models exhibited severe loss-wise overfitting with loss ratios reaching 
$10.16$ for CIFAR-100 and $71.0$ for CIFAR-10, despite a small test accuracy gap for CIFAR-10. 
In contrast, AOF and TL dramatically reduced these ratios to below 2,
which means that there is no contradiction between utility and reducing vulnerability to MIAs. 
In our experiments, the loss ratio showed a strong correlation with the vulnerability to LiRA (see Section~\ref{sec:lratio}).
In particular, the accuracy gaps alone failed to capture the magnitude of the vulnerability because it is insensitive to the confidence distribution. 
Models can achieve similar test accuracy while exhibiting vastly different loss ratios and the corresponding MIA 
vulnerability. 
Appendix~\ref{app:ablation} presents detailed ablations on individual anti-overfitting components and their optimized combinations.
\begin{table}[t!]
\centering
\caption{Model utility and loss ratio across benchmarks.}
\label{tab:utility}
\resizebox{\linewidth}{!}{
\begin{tabular}{@{}lcccccc@{}} 
\toprule
Benchmark & Train Loss & Test Loss & Loss Ratio & Train Acc (\%) & Test Acc (\%) \\ 
\midrule
CIFAR-10 (baseline) & 0.0032 & 0.2272 & 71.0 & 99.94 & 93.63 \\
CIFAR-10 (AOF) & 0.1351 & 0.2535 & 1.88 & 98.78 & 94.09 \\
CIFAR-10 (AOF+TL) & 0.1210 & 0.1647 & 1.36 & 98.70 & 97.00 \\ 
\cmidrule(r){1-6}
CIFAR-100 (baseline) & 0.1198 & 1.2172 & 10.16 & 96.75 & 69.30 \\
CIFAR-100 (AOF) & 0.8007 & 1.2037 & 1.50 & 79.75 & 68.24 \\
CIFAR-100 (AOF+TL) & 0.3373 & 0.6214 & 1.84 & 92.71 & 83.56 \\ 
\cmidrule(r){1-6}
GTSRB (baseline) & 0.0448 & 0.0614 & 1.37 & 99.30 & 98.69 \\
GTSRB (AOF+TL) & 0.3145 & 0.3175 & 1.01 & 99.29 & 98.94 \\ 
\cmidrule(r){1-6}
Purchase-100 (baseline) & 0.1096 & 0.1861 & 1.70 & 99.70 & 94.91 \\
Purchase-100 (AOF) & 0.2266 & 0.2763 & 1.22 & 96.66 & 93.18 \\
\bottomrule
\end{tabular}
}
\end{table}

\paragraph{Attack effectiveness under optimistic evaluation}
To isolate the effects of AOF and TL, 
we first evaluated under target-calibrated thresholds and balanced prior ($\pi=50\%$), 
which represents the upper bounds of MIA risk. 
Tables~\ref{tab:cifar10-optm}--\ref{tab:purchase-optm} present the results. 
For datasets with high baseline loss ratios (CIFAR-10 and CIFAR-100), 
all LiRA variants initially achieved high TPR and AUC. 
In contrast, benchmarks with naturally low loss ratios (GTSRB with $1.37$ and Purchase-100 
with $1.70$) exhibited minimal attack success, especially the well-generalized GTSRB.
The introduction of AOF dramatically reduced the attack effectiveness for all benchmarks, especially those with a high baseline loss ratio. 
For CIFAR-10, the strongest online LiRA variant dropped from $10.268\%$ to $2.723\%$ TPR at FPR=$0.1\%$ (a $3.8\times$
reduction) and from $3.956\%$ to $0.248\%$ at FPR=$0.001\%$ (a $16\times$ reduction). 
Adding TL compounded these reductions: the same attack endpoints fell
to $0.521\%$ and $0.065\%$, respectively, representing $20\times$ and $61\times$ reductions from baseline.
Similar patterns emerged across all datasets, demonstrating that privacy protection does not need to compromise (and can actually enhance) the utility of the model.
The offline LiRA variants, which rely on one-sided hypothesis testing, suffered even
worse once overfitting was controlled and in most cases approached random guessing (AUC$\approx 50\%$).  
Interestingly, fixed-variance (FV) variants showed marginally better stability
than per-sample variance estimation after AOF/TL application, 
suggesting that per-sample variance estimation becomes unreliable when
models generalize well.
We can also observe that AUC poorly reflects privacy risk in the ultra-low-FPR regime
critical for practical deployment. 
For example, for the CIFAR-10 baseline, the simple global threshold
attack achieved higher AUC than the offline variants, yet yielded TPR far below the latter. 
This confirms previous observations that relying on average metrics, such as accuracy or AUC, can be misleading~\cite{ye2022enhanced,carlini2022membership}.

In summary, across all benchmarks and LiRA variants, the reductions in TPR with AOF ranged from $2.4\times$ 
to $18\times$ with an average of $6.2\times$.
TL amplified these reductions to $191\times$ with an average of $28\times$. 
These dramatic reductions under optimistic evaluation conditions demonstrate 
that \textbf{properly tuned anti-overfitting techniques and transfer learning can cut
most of LiRA's effectiveness by addressing the root cause: loss-wise overfitting that creates
exploitable confidence disparities between member and non-member samples}. 
\begin{table}[t!]
\centering
\caption{CIFAR-10 with proper AOF and TL. TPR reduction factors relative to baseline
shown in parentheses for AOF and TL. Standard deviations reported in small font.}
\label{tab:cifar10-optm}
\small
\resizebox{\linewidth}{!}{
\begin{tabular}{lcccc}
\toprule
Benchmark & Attack & TPR@0.001\% FPR (\%)& TPR@0.1\% FPR (\%)& AUC (\%)\\ 
\midrule
\multirow{5}{*}{Baseline} 
& Online & 3.956$_{\pm 1.061}$ & 10.268$_{\pm 0.555}$ & 76.48$_{\pm 0.32}$ \\
& Online (FV) & 2.876$_{\pm 1.064}$ & 9.135$_{\pm 0.508}$ & 76.28$_{\pm 0.31}$ \\
& Offline & 0.762$_{\pm 0.348}$ & 3.262$_{\pm 0.338}$ & 55.58$_{\pm 0.92}$ \\
& Offline (FV) & 0.948$_{\pm 0.526}$ & 4.540$_{\pm 0.424}$ & 56.64$_{\pm 0.89}$ \\
& Global & 0.003$_{\pm 0.004}$ & 0.097$_{\pm 0.027}$ & 59.97$_{\pm 0.32}$ \\
\cmidrule(lr){1-5}
\multirow{5}{*}{AOF} 
& Online & 0.248$_{\pm 0.198}$ ($\times$16) & 2.723$_{\pm 0.683}$ ($\times$3.8) & 65.76$_{\pm 1.27}$ \\
& Online (FV) & 0.687$_{\pm 0.351}$ ($\times$4.2) & 3.483$_{\pm 0.405}$ ($\times$2.6) & 68.26$_{\pm 1.57}$ \\
& Offline & 0.042$_{\pm 0.036}$ ($\times$18) & 0.539$_{\pm 0.132}$ ($\times$6.1) & 49.31$_{\pm 1.73}$ \\
& Offline (FV) & 0.254$_{\pm 0.147}$ ($\times$3.7) & 1.479$_{\pm 0.222}$ ($\times$3.1) & 48.99$_{\pm 2.35}$ \\
& Global & 0.003$_{\pm 0.005}$ ($\times$1.0) & 0.110$_{\pm 0.028}$ ($\times$0.9) & 56.19$_{\pm 0.50}$ \\
\cmidrule(lr){1-5}
\multirow{5}{*}{AOF+TL} 
& Online & 0.065$_{\pm 0.061}$ ($\times$61) & 0.521$_{\pm 0.128}$ ($\times$20) & 56.57$_{\pm 0.36}$ \\
& Online (FV) & 0.092$_{\pm 0.058}$ ($\times$31) & 0.834$_{\pm 0.106}$ ($\times$11) & 57.07$_{\pm 0.36}$ \\
& Offline & 0.004$_{\pm 0.005}$ ($\times$190) & 0.097$_{\pm 0.028}$ ($\times$34) & 49.92$_{\pm 1.07}$ \\
& Offline (FV) & 0.016$_{\pm 0.020}$ ($\times$59) & 0.253$_{\pm 0.079}$ ($\times$18) & 50.03$_{\pm 1.22}$ \\
& Global & 0.005$_{\pm 0.006}$ ($\times$0.6) & 0.114$_{\pm 0.027}$ ($\times$0.9) & 53.06$_{\pm 0.26}$ \\
\bottomrule
\end{tabular}
}
\end{table}

\begin{table}[t!]
\centering
\caption{CIFAR-100 with proper AOF and TL.}
\label{tab:cifar100-optm}
\small
\resizebox{\linewidth}{!}{
\begin{tabular}{lcccc}
\toprule
Benchmark & Attack & TPR@0.001\% FPR (\%)& TPR@0.1\% FPR (\%)& AUC (\%)\\ 
\midrule
\multirow{5}{*}{Baseline} 
& Online & 4.619$_{\pm 1.730}$ & 15.791$_{\pm 0.976}$ & 88.28$_{\pm 0.21}$ \\
& Online (FV) & 1.730$_{\pm 1.158}$ & 11.306$_{\pm 1.000}$ & 87.81$_{\pm 0.21}$ \\
& Offline & 0.659$_{\pm 0.502}$ & 6.044$_{\pm 0.640}$ & 72.10$_{\pm 0.40}$ \\
& Offline (FV) & 0.253$_{\pm 0.234}$ & 3.911$_{\pm 0.524}$ & 71.88$_{\pm 0.39}$ \\
& Global & 0.002$_{\pm 0.004}$ & 0.098$_{\pm 0.026}$ & 69.86$_{\pm 0.28}$ \\
\cmidrule(lr){1-5}
\multirow{5}{*}{AOF} 
& Online & 0.351$_{\pm 0.226}$ ($\times$13) & 3.097$_{\pm 0.353}$ ($\times$5.1) & 75.32$_{\pm 0.31}$ \\
& Online (FV) & 0.214$_{\pm 0.162}$ ($\times$8.1) & 2.404$_{\pm 0.300}$ ($\times$4.7) & 74.96$_{\pm 0.32}$ \\
& Offline & 0.090$_{\pm 0.069}$ ($\times$7.3) & 1.141$_{\pm 0.205}$ ($\times$5.3) & 58.25$_{\pm 0.84}$ \\
& Offline (FV) & 0.105$_{\pm 0.086}$ ($\times$2.4) & 1.240$_{\pm 0.193}$ ($\times$3.2) & 58.81$_{\pm 0.81}$ \\
& Global & 0.004$_{\pm 0.006}$ ($\times$0.5) & 0.155$_{\pm 0.035}$ ($\times$0.6) & 58.82$_{\pm 0.24}$ \\
\cmidrule(lr){1-5}
\multirow{5}{*}{AOF+TL} 
& Online & 0.270$_{\pm 0.237}$ ($\times$17) & 2.367$_{\pm 0.398}$ ($\times$6.7) & 67.82$_{\pm 0.33}$ \\
& Online (FV) & 0.243$_{\pm 0.183}$ ($\times$7.1) & 2.223$_{\pm 0.265}$ ($\times$5.1) & 68.07$_{\pm 0.38}$ \\
& Offline & 0.012$_{\pm 0.013}$ ($\times$55) & 0.294$_{\pm 0.087}$ ($\times$21) & 52.56$_{\pm 0.98}$ \\
& Offline (FV) & 0.046$_{\pm 0.045}$ ($\times$5.5) & 0.734$_{\pm 0.179}$ ($\times$5.3) & 53.50$_{\pm 1.02}$ \\
& Global & 0.006$_{\pm 0.007}$ ($\times$0.3) & 0.158$_{\pm 0.035}$ ($\times$0.6) & 58.47$_{\pm 0.26}$ \\
\bottomrule
\end{tabular}
}
\end{table}

\begin{table}[t!]
\centering
\caption{GTSRB with proper AOF and TL.}
\label{tab:gtsrb-optm}
\small
\resizebox{\linewidth}{!}{%
\begin{tabular}{lcccc}
\toprule
Benchmark & Attack & TPR@0.001\% FPR (\%)& TPR@0.1\% FPR (\%)& AUC (\%)\\ 
\midrule
\multirow{5}{*}{Baseline} 
& Online & 0.039$_{\pm 0.028}$ & 0.274$_{\pm 0.059}$ & 53.09$_{\pm 0.27}$ \\
& Online (FV) & 0.042$_{\pm 0.032}$ & 0.319$_{\pm 0.068}$ & 53.10$_{\pm 0.28}$ \\
& Offline & 0.002$_{\pm 0.004}$ & 0.064$_{\pm 0.034}$ & 49.47$_{\pm 0.62}$ \\
& Offline (FV) & 0.005$_{\pm 0.007}$ & 0.085$_{\pm 0.038}$ & 49.44$_{\pm 0.63}$ \\
& Global & 0.006$_{\pm 0.008}$ & 0.100$_{\pm 0.033}$ & 51.10$_{\pm 0.29}$ \\
\cmidrule(lr){1-5}
\multirow{5}{*}{AOF+TL} 
& Online & 0.006$_{\pm 0.007}$ ($\times$6.5) & 0.109$_{\pm 0.033}$ ($\times$2.5) & 51.68$_{\pm 0.40}$ \\
& Online (FV) & 0.019$_{\pm 0.017}$ ($\times$2.2) & 0.225$_{\pm 0.051}$ ($\times$1.4) & 52.11$_{\pm 0.40}$ \\
& Offline & 0.005$_{\pm 0.008}$ ($\times$0.4) & 0.094$_{\pm 0.036}$ ($\times$0.7) & 49.82$_{\pm 0.72}$ \\
& Offline (FV) & 0.006$_{\pm 0.008}$ ($\times$0.8) & 0.104$_{\pm 0.041}$ ($\times$0.8) & 49.79$_{\pm 0.77}$ \\
& Global & 0.007$_{\pm 0.009}$ ($\times$0.9) & 0.126$_{\pm 0.035}$ ($\times$0.8) & 51.62$_{\pm 0.33}$ \\
\bottomrule
\end{tabular}%
}
\end{table}

\begin{table}[t!]
\centering
\caption{Purchase-100 with proper AOF.}
\label{tab:purchase-optm}
\small
\resizebox{\linewidth}{!}{%
\begin{tabular}{lcccc}
\toprule
Benchmark & Attack & TPR@0.001\% FPR (\%)& TPR@0.1\% FPR (\%)& AUC (\%)\\ 
\midrule
\multirow{5}{*}{Baseline} 
& Online & 0.523$_{\pm 0.243}$ & 4.491$_{\pm 0.281}$ & 70.16$_{\pm 0.29}$ \\
& Online (FV) & 0.180$_{\pm 0.110}$ & 3.089$_{\pm 0.188}$ & 69.52$_{\pm 0.28}$ \\
& Offline & 0.007$_{\pm 0.007}$ & 0.500$_{\pm 0.077}$ & 55.11$_{\pm 0.48}$ \\
& Offline (FV) & 0.022$_{\pm 0.017}$ & 0.713$_{\pm 0.078}$ & 56.11$_{\pm 0.51}$ \\
& Global & 0.001$_{\pm 0.001}$ & 0.100$_{\pm 0.015}$ & 54.83$_{\pm 0.15}$ \\
\cmidrule(lr){1-5}
\multirow{5}{*}{AOF} 
& Online & 0.022$_{\pm 0.017}$ ($\times$24) & 0.825$_{\pm 0.068}$ ($\times$5.4) & 62.64$_{\pm 0.16}$ \\
& Online (FV) & 0.026$_{\pm 0.019}$ ($\times$6.9) & 0.794$_{\pm 0.067}$ ($\times$3.9) & 62.82$_{\pm 0.16}$ \\
& Offline & 0.001$_{\pm 0.001}$ ($\times$7.0) & 0.139$_{\pm 0.022}$ ($\times$3.6) & 51.88$_{\pm 0.18}$ \\
& Offline (FV) & 0.003$_{\pm 0.003}$ ($\times$7.3) & 0.230$_{\pm 0.031}$ ($\times$3.1) & 52.50$_{\pm 0.19}$ \\
& Global & 0.001$_{\pm 0.002}$ ($\times$1.0) & 0.101$_{\pm 0.014}$ ($\times$1.0) & 52.20$_{\pm 0.13}$ \\
\bottomrule
\end{tabular}%
}
\end{table}

\subsection{Impact of Skewed Priors and Shadow-based Thresholds}
\label{sec:realistic}
After establishing that AOF and TL substantially reduce the effectiveness of LiRA, 
we now examine whether any residual success translates into \emph{reliable} membership inferences (\emph{i.e.}, high-certainty positives)
under \emph{shadow-based} thresholds and \emph{skewed} priors.
Tables~\ref{tab:cifar10_both}--\ref{tab:purchase100_both} report results at a nominal target FPR of $0.001\%$, contrasting an \emph{optimistic} setting (target-calibrated threshold, $\pi{=}50\%$) with a \emph{realistic} setting (shadow-calibrated threshold, $\pi \in \{1\%,10\%,50\%\}$).
In the optimistic setting, the false positive rate FPR$'$ achieved among all variants was $0$, which yielded perfect PPV of 100\% for any prior $\pi$ due to threshold overfitting to the target.

In the realistic setting using \emph{shadow-based} thresholding, the results changed substantially.
In the \emph{overfitted baselines} (CIFAR-10/100), PPVs were not markedly affected across priors because overfitting dominates the signal.
Yet FPR$'$ became non-zero and exhibited variability.
More importantly, once AOF (and especially AOF+TL) were applied, we observed
(i) significantly increased and more variable FPR$'$ and 
(ii) slight to modest change (mostly decreased) in the achieved TPR$'$.
For example, on CIFAR-10 with AOF, online LiRA’s FPR$'$ increased to $0.033\%\pm0.466\%$ (std exceeding the mean), which decreased PPV to $90.93\%\pm12.18\%$ at $\pi{=}10\%$ and to $66.52\%\pm34.88\%$ at $\pi{=}1\%$.
These changes with the corresponding high-variance indicate limited transferability of thresholds when models generalize well.
Adding TL further degraded PPV (and increased variance) due to the combined effect of slightly lower TPR$'$ and higher FPR$'$. 
For instance, on CIFAR-10 with AOF+TL, online LiRA’s PPV at $\pi{=}10\%$ dropped to $70.75\%\pm30.40\%$.
Less expensive offline variants were worse and their PPVs frequently fell to substantially less reliable levels across benchmarks under AOF/AOF+TL, particularly for $\pi \le 10\%$.
This is because AOF/TL shrank the separation between IN and OUT logit distributions, causing overlap and, therefore, deviating thresholds from the optimal ones.
PPV’s sensitivity to prior was driven primarily by the achieved FPR$'$: the larger (and more variable) the FPR$'$, the stronger the decrease in PPV as $\pi$ decreased. 
Note that skewed priors do not change the ROC trade-off and are not unique to MIAs, but assuming $\pi{=}50\%$ can misrepresent MIA risk~\cite{jayaraman2021revisiting}. 
Although PPV is prior-dependent, our results show that its degradation was primarily driven by the interaction between skewed priors and miscalibrated FPR$'$.
Across variants, fixed-variance (FV) LiRA often achieved lower FPR$'$ and PPV with lower standard deviations than per-sample variance, suggesting that a global variance estimate is more robust once the overfitting is reduced.
Models that met deployment standards (\emph{e.g.}, GTSRB) yielded very low attack effectiveness under realistic calibration and priors, with almost no reliable membership signal.\\

From these observations, two implications emerge:
\paragraph{(1) Poor threshold transferability constrains attack reliability}
Even a well-resourced attacker cannot reliably calibrate the thresholds without access to the \emph{target’s} score distribution.
The thresholds learned from shadows---the only realistic option in black-box settings---deviated from the target FPR, 
producing high variance among the FPR$'$, TPR$'$, and PPV achieved for the targets.
This is because AOF and TL shrink the member vs.\ non-member confidence distributions (see Fig.~\ref{fig:sample_inout_dist}), and when combined with training stochasticity, 
models converge to different local minima with distinct calibration.
As a result, likelihood-ratio tails become more model-dependent, causing the locally obtained ``optimal'' threshold to be \emph{less transferable}.

\paragraph{(2) Imperfect precision reduces the reliability of individual inferences}
Under realistic priors ($\pi \le 10\%$) and shadow calibration, 
LiRA's PPV frequently fell well below the near-perfect levels required for confident subject-level claims.
With AOF, PPV often decreased to 80--90\% in $\pi{=}10\%$ and 60--70\% in $\pi{=}1\%$. 
Adding TL further degraded precision, sometimes to 25--50\% in the weakest cases.
At these levels, a substantial fraction of flagged samples are false positives, 
granting individuals considerable \emph{plausible deniability}~\cite{bindschaedler2017plausible} 
due to the increased uncertainty in positive inferences.
This uncertainty also increases
as the membership prior decreases, which is the relevant regime when targeting specific individuals in sensitive domains.
We do not claim that moderate PPV eliminates privacy risk, especially under skewed priors where it may still provide targets for further investigation. 
Rather, lower PPV weakens the evidentiary strength of individual membership claims.

Together, these findings indicate that \textbf{under realistic calibration and priors, LiRA's residual success often translates into substantially higher uncertainty in positive inferences
rather than actionable privacy leakage.}
\begin{table}[t!]
\caption{CIFAR-10 under target vs.\ shadow calibration (target FPR $=0.001\%$).}
\label{tab:cifar10_both}
\centering
\small
\resizebox{\linewidth}{!}{
\begin{tabular}{ll|cc|ccc}
\toprule
 & & \multicolumn{2}{c|}{Performance} & \multicolumn{3}{c}{PPV (\%)} \\
Benchmark & Attack & TPR$'$ (\%)& FPR$'$ (\%)& @$\pi$=1\% & @$\pi$=10\% & @$\pi$=50\% \\
\midrule
\multicolumn{7}{l}{\textit{\textbf{Target-based thresholds (optimistic)}}} \\
\cmidrule(lr){1-7}
\multirow{4}{*}{Baseline} 
& Online       & 3.956$_{\pm 1.061}$ & 0.000$_{\pm 0.000}$ & 100.00$_{\pm 0.00}$ & 100.00$_{\pm 0.00}$ & 100.00$_{\pm 0.00}$ \\
& Online (FV)  & 2.876$_{\pm 1.064}$ & 0.000$_{\pm 0.000}$ & 100.00$_{\pm 0.00}$ & 100.00$_{\pm 0.00}$ & 100.00$_{\pm 0.00}$ \\
& Offline      & 0.762$_{\pm 0.348}$ & 0.000$_{\pm 0.000}$ & 100.00$_{\pm 0.00}$ & 100.00$_{\pm 0.00}$ & 100.00$_{\pm 0.00}$ \\
& Offline (FV) & 0.948$_{\pm 0.526}$ & 0.000$_{\pm 0.000}$ & 100.00$_{\pm 0.00}$ & 100.00$_{\pm 0.00}$ & 100.00$_{\pm 0.00}$ \\
\cmidrule(lr){1-7}
\multicolumn{7}{l}{\textit{\textbf{Shadow-based thresholds (realistic)}}} \\
\cmidrule(lr){1-7}
\multirow{4}{*}{Baseline} 
& Online       & 3.990$_{\pm 0.161}$ & 0.002$_{\pm 0.003}$ & 94.73$_{\pm 6.10}$ & 99.46$_{\pm 0.65}$ & 99.94$_{\pm 0.07}$ \\
& Online (FV)  & 2.912$_{\pm 0.142}$ & 0.002$_{\pm 0.003}$ & 93.10$_{\pm 8.03}$ & 99.26$_{\pm 0.91}$ & 99.92$_{\pm 0.10}$ \\
& Offline      & 0.713$_{\pm 0.052}$ & 0.002$_{\pm 0.003}$ & 81.31$_{\pm 20.20}$ & 97.24$_{\pm 3.33}$ & 99.67$_{\pm 0.40}$ \\
& Offline (FV) & 0.918$_{\pm 0.068}$ & 0.003$_{\pm 0.005}$ & 81.13$_{\pm 21.29}$ & 97.03$_{\pm 4.06}$ & 99.64$_{\pm 0.52}$ \\
\cmidrule(lr){2-7}
\multirow{4}{*}{AOF} 
& Online       & 0.224$_{\pm 0.482}$ & 0.033$_{\pm 0.466}$ & 66.52$_{\pm 34.88}$ & 90.93$_{\pm 12.18}$ & 98.42$_{\pm 5.25}$ \\
& Online (FV)  & 0.636$_{\pm 0.101}$ & 0.002$_{\pm 0.003}$ & 80.13$_{\pm 21.52}$ & 96.69$_{\pm 6.53}$ & 99.46$_{\pm 2.99}$ \\
& Offline      & 0.290$_{\pm 4.134}$ & 0.262$_{\pm 4.138}$ & 55.17$_{\pm 46.40}$ & 73.31$_{\pm 28.84}$ & 93.37$_{\pm 9.32}$ \\
& Offline (FV) & 0.310$_{\pm 1.179}$ & 0.077$_{\pm 1.192}$ & 67.96$_{\pm 34.53}$ & 91.18$_{\pm 12.71}$ & 98.36$_{\pm 6.23}$ \\
\cmidrule(lr){2-7}
\multirow{4}{*}{AOF+TL} 
& Online       & 0.084$_{\pm 0.048}$ & 0.017$_{\pm 0.045}$ & 49.13$_{\pm 44.90}$ & 70.75$_{\pm 30.40}$ & 91.49$_{\pm 12.35}$ \\
& Online (FV)  & 0.084$_{\pm 0.021}$ & 0.002$_{\pm 0.003}$ & 59.25$_{\pm 42.01}$ & 83.54$_{\pm 18.38}$ & 97.22$_{\pm 3.42}$ \\
& Offline      & 0.027$_{\pm 0.085}$ & 0.026$_{\pm 0.085}$ & 32.73$_{\pm 46.50}$ & 37.30$_{\pm 43.64}$ & 56.17$_{\pm 36.15}$ \\
& Offline (FV) & 0.044$_{\pm 0.089}$ & 0.033$_{\pm 0.089}$ & 42.39$_{\pm 48.08}$ & 52.67$_{\pm 40.53}$ & 78.43$_{\pm 21.99}$ \\
\bottomrule
\end{tabular}}
\end{table}

\begin{table}[t!]
\caption{CIFAR-100 under target vs.\ shadow calibration (target FPR $=0.001\%$).}
\label{tab:cifar100_both}
\centering
\small
\resizebox{\linewidth}{!}{
\begin{tabular}{ll|cc|ccc}
\toprule
 & & \multicolumn{2}{c|}{Performance} & \multicolumn{3}{c}{PPV (\%)} \\
Benchmark & Attack & TPR$'$ (\%)& FPR$'$ (\%)& @$\pi$=1\% & @$\pi$=10\% & @$\pi$=50\% \\
\midrule
\multicolumn{7}{l}{\textit{\textbf{Target-based thresholds (optimistic)}}} \\
\cmidrule(lr){1-7}
\multirow{4}{*}{Baseline} 
& Online       & 4.619$_{\pm 1.730}$ & 0.000$_{\pm 0.000}$ & 100.00$_{\pm 0.00}$ & 100.00$_{\pm 0.00}$ & 100.00$_{\pm 0.00}$ \\
& Online (FV)  & 1.730$_{\pm 1.158}$ & 0.000$_{\pm 0.000}$ & 100.00$_{\pm 0.00}$ & 100.00$_{\pm 0.00}$ & 100.00$_{\pm 0.00}$ \\
& Offline      & 0.659$_{\pm 0.502}$ & 0.000$_{\pm 0.000}$ & 100.00$_{\pm 0.00}$ & 100.00$_{\pm 0.00}$ & 100.00$_{\pm 0.00}$ \\
& Offline (FV) & 0.253$_{\pm 0.234}$ & 0.000$_{\pm 0.000}$ & 100.00$_{\pm 0.00}$ & 100.00$_{\pm 0.00}$ & 100.00$_{\pm 0.00}$ \\
\cmidrule(lr){1-7}
\multicolumn{7}{l}{\textit{\textbf{Shadow-based thresholds (realistic)}}} \\
\cmidrule(lr){1-7}
\multirow{4}{*}{Baseline} 
& Online       & 4.670$_{\pm 0.197}$ & 0.003$_{\pm 0.003}$ & 95.19$_{\pm 5.67}$ & 99.51$_{\pm 0.60}$ & 99.95$_{\pm 0.07}$ \\
& Online (FV)  & 1.595$_{\pm 0.097}$ & 0.002$_{\pm 0.003}$ & 88.86$_{\pm 12.36}$ & 98.68$_{\pm 1.57}$ & 99.85$_{\pm 0.18}$ \\
& Offline      & 0.555$_{\pm 0.057}$ & 0.002$_{\pm 0.003}$ & 78.69$_{\pm 22.43}$ & 96.64$_{\pm 3.96}$ & 99.60$_{\pm 0.49}$ \\
& Offline (FV) & 0.196$_{\pm 0.030}$ & 0.003$_{\pm 0.003}$ & 65.09$_{\pm 35.23}$ & 90.68$_{\pm 10.65}$ & 98.71$_{\pm 1.63}$ \\
\cmidrule(lr){2-7}
\multirow{4}{*}{AOF} 
& Online       & 0.296$_{\pm 0.042}$ & 0.002$_{\pm 0.003}$ & 70.36$_{\pm 30.45}$ & 93.69$_{\pm 7.29}$ & 99.19$_{\pm 0.98}$ \\
& Online (FV)  & 0.170$_{\pm 0.027}$ & 0.002$_{\pm 0.003}$ & 65.70$_{\pm 35.60}$ & 90.71$_{\pm 10.78}$ & 98.71$_{\pm 1.64}$ \\
& Offline      & 0.070$_{\pm 0.018}$ & 0.002$_{\pm 0.003}$ & 58.18$_{\pm 42.73}$ & 82.14$_{\pm 19.85}$ & 96.81$_{\pm 4.18}$ \\
& Offline (FV) & 0.076$_{\pm 0.019}$ & 0.002$_{\pm 0.003}$ & 59.09$_{\pm 42.49}$ & 82.72$_{\pm 19.71}$ & 96.91$_{\pm 4.14}$ \\
\cmidrule(lr){2-7}
\multirow{4}{*}{AOF+TL} 
& Online       & 0.240$_{\pm 0.038}$ & 0.006$_{\pm 0.018}$ & 64.57$_{\pm 35.05}$ & 89.52$_{\pm 14.82}$ & 98.19$_{\pm 3.94}$ \\
& Online (FV)  & 0.204$_{\pm 0.036}$ & 0.002$_{\pm 0.002}$ & 68.38$_{\pm 32.86}$ & 92.61$_{\pm 8.38}$ & 99.03$_{\pm 1.15}$ \\
& Offline      & 0.018$_{\pm 0.049}$ & 0.011$_{\pm 0.047}$ & 46.06$_{\pm 48.94}$ & 53.92$_{\pm 42.42}$ & 76.41$_{\pm 26.52}$ \\
& Offline (FV) & 0.049$_{\pm 0.049}$ & 0.012$_{\pm 0.049}$ & 50.43$_{\pm 46.40}$ & 69.37$_{\pm 31.01}$ & 91.38$_{\pm 11.63}$ \\
\bottomrule
\end{tabular}}
\end{table}

\begin{table}[t!]
\caption{GTSRB under target vs.\ shadow calibration (target FPR $=0.001\%$).}
\label{tab:gtsrb_both}
\centering
\small
\resizebox{\linewidth}{!}{
\begin{tabular}{ll|cc|ccc}
\toprule
 & & \multicolumn{2}{c|}{Performance} & \multicolumn{3}{c}{PPV (\%)} \\
Benchmark & Attack & TPR$'$ (\%)& FPR$'$ (\%)& @$\pi$=1\% & @$\pi$=10\% & @$\pi$=50\% \\
\midrule
\multicolumn{7}{l}{\textit{\textbf{Target-based thresholds (optimistic)}}} \\
\cmidrule(lr){1-7}
\multirow{4}{*}{Baseline} 
& Online       & 0.039$_{\pm 0.028}$ & 0.000$_{\pm 0.000}$ & 100.00$_{\pm 0.00}$ & 100.00$_{\pm 0.00}$ & 100.00$_{\pm 0.00}$ \\
& Online (FV)  & 0.042$_{\pm 0.032}$ & 0.000$_{\pm 0.000}$ & 100.00$_{\pm 0.00}$ & 100.00$_{\pm 0.00}$ & 100.00$_{\pm 0.00}$ \\
& Offline      & 0.002$_{\pm 0.004}$ & 0.000$_{\pm 0.000}$ & 100.00$_{\pm 0.00}$ & 100.00$_{\pm 0.00}$ & 100.00$_{\pm 0.00}$ \\
& Offline (FV) & 0.005$_{\pm 0.007}$ & 0.000$_{\pm 0.000}$ & 100.00$_{\pm 0.00}$ & 100.00$_{\pm 0.00}$ & 100.00$_{\pm 0.00}$ \\
\cmidrule(lr){1-7}
\multicolumn{7}{l}{\textit{\textbf{Shadow-based thresholds (realistic)}}} \\
\cmidrule(lr){1-7}
\multirow{4}{*}{Baseline} 
& Online       & 0.031$_{\pm 0.013}$ & 0.003$_{\pm 0.005}$ & 59.13$_{\pm 47.20}$ & 71.83$_{\pm 33.34}$ & 91.61$_{\pm 11.37}$ \\
& Online (FV)  & 0.035$_{\pm 0.013}$ & 0.003$_{\pm 0.005}$ & 57.13$_{\pm 47.22}$ & 71.27$_{\pm 32.66}$ & 91.67$_{\pm 11.23}$ \\
& Offline      & 0.003$_{\pm 0.006}$ & 0.007$_{\pm 0.011}$ & 11.57$_{\pm 31.58}$ & 14.29$_{\pm 31.18}$ & 24.85$_{\pm 34.80}$ \\
& Offline (FV) & 0.004$_{\pm 0.005}$ & 0.005$_{\pm 0.008}$ & 35.13$_{\pm 47.57}$ & 37.59$_{\pm 46.02}$ & 47.72$_{\pm 43.17}$ \\
\cmidrule(lr){2-7}
\multirow{4}{*}{AOF+TL} 
& Online       & 0.038$_{\pm 0.109}$ & 0.037$_{\pm 0.106}$ & 26.21$_{\pm 43.27}$ & 32.50$_{\pm 40.06}$ & 56.85$_{\pm 31.44}$ \\
& Online (FV)  & 0.017$_{\pm 0.019}$ & 0.005$_{\pm 0.017}$ & 53.83$_{\pm 48.80}$ & 62.45$_{\pm 40.28}$ & 83.98$_{\pm 19.61}$ \\
& Offline      & 0.071$_{\pm 0.219}$ & 0.069$_{\pm 0.215}$ & 15.93$_{\pm 35.70}$ & 22.29$_{\pm 33.59}$ & 47.10$_{\pm 32.04}$ \\
& Offline (FV) & 0.080$_{\pm 0.219}$ & 0.079$_{\pm 0.221}$ & 18.00$_{\pm 37.39}$ & 25.24$_{\pm 34.83}$ & 51.88$_{\pm 30.75}$ \\
\bottomrule
\end{tabular}}
\end{table}

\begin{table}[t!]
\caption{Purchase-100 under target vs.\ shadow calibration (target FPR $=0.001\%$).}
\label{tab:purchase100_both}
\centering
\small
\resizebox{\linewidth}{!}{
\begin{tabular}{ll|cc|ccc}
\toprule
 & & \multicolumn{2}{c|}{Performance} & \multicolumn{3}{c}{PPV (\%)} \\
Benchmark & Attack & TPR$'$ (\%)& FPR$'$ (\%)& @$\pi$=1\% & @$\pi$=10\% & @$\pi$=50\% \\
\midrule
\multicolumn{7}{l}{\textit{\textbf{Target-based thresholds (optimistic)}}} \\
\cmidrule(lr){1-7}
\multirow{4}{*}{Baseline} 
& Online       & 0.523$_{\pm 0.243}$ & 0.000$_{\pm 0.000}$ & 100.00$_{\pm 0.00}$ & 100.00$_{\pm 0.00}$ & 100.00$_{\pm 0.00}$ \\
& Online (FV)  & 0.180$_{\pm 0.110}$ & 0.000$_{\pm 0.000}$ & 100.00$_{\pm 0.00}$ & 100.00$_{\pm 0.00}$ & 100.00$_{\pm 0.00}$ \\
& Offline      & 0.007$_{\pm 0.007}$ & 0.000$_{\pm 0.000}$ & 100.00$_{\pm 0.00}$ & 100.00$_{\pm 0.00}$ & 100.00$_{\pm 0.00}$ \\
& Offline (FV) & 0.022$_{\pm 0.017}$ & 0.000$_{\pm 0.000}$ & 100.00$_{\pm 0.00}$ & 100.00$_{\pm 0.00}$ & 100.00$_{\pm 0.00}$ \\
\cmidrule(lr){1-7}
\multicolumn{7}{l}{\textit{\textbf{Shadow-based thresholds (realistic)}}} \\
\cmidrule(lr){1-7}
\multirow{4}{*}{Baseline} 
& Online       & 0.516$_{\pm 0.047}$ & 0.001$_{\pm 0.001}$ & 89.93$_{\pm 11.15}$ & 98.84$_{\pm 1.37}$ & 99.87$_{\pm 0.16}$ \\
& Online (FV)  & 0.159$_{\pm 0.015}$ & 0.001$_{\pm 0.001}$ & 78.87$_{\pm 22.27}$ & 96.70$_{\pm 3.80}$ & 99.61$_{\pm 0.47}$ \\
& Offline      & 0.004$_{\pm 0.002}$ & 0.001$_{\pm 0.001}$ & 55.82$_{\pm 48.27}$ & 66.20$_{\pm 37.65}$ & 87.37$_{\pm 16.56}$ \\
& Offline (FV) & 0.018$_{\pm 0.004}$ & 0.001$_{\pm 0.001}$ & 57.27$_{\pm 43.60}$ & 80.46$_{\pm 21.53}$ & 96.32$_{\pm 4.79}$ \\
\cmidrule(lr){2-7}
\multirow{4}{*}{AOF} 
& Online       & 0.019$_{\pm 0.005}$ & 0.001$_{\pm 0.001}$ & 57.73$_{\pm 43.46}$ & 81.14$_{\pm 20.47}$ & 96.63$_{\pm 4.01}$ \\
& Online (FV)  & 0.022$_{\pm 0.005}$ & 0.001$_{\pm 0.001}$ & 58.46$_{\pm 42.42}$ & 82.88$_{\pm 18.74}$ & 97.07$_{\pm 3.57}$ \\
& Offline      & 0.001$_{\pm 0.001}$ & 0.001$_{\pm 0.001}$ & 27.29$_{\pm 44.29}$ & 30.20$_{\pm 42.78}$ & 42.92$_{\pm 40.47}$ \\
& Offline (FV) & 0.002$_{\pm 0.001}$ & 0.001$_{\pm 0.001}$ & 44.22$_{\pm 48.77}$ & 51.95$_{\pm 42.69}$ & 73.99$_{\pm 29.04}$ \\
\bottomrule
\end{tabular}}
\end{table}

\subsection{Reproducibility}
\label{sec:reproducibility}
When LiRA is used to select a small subset of training samples for targeted protection or as ground-truth labels for evaluating efficient alternatives~\cite{jiacheng2024mist,pollock2025free}, a key question is how reproducible the resulting thresholded set is under run-to-run variability.
Substantial cross-run variation implies that a single-run thresholded set may fail to reliably capture all high-risk samples.
We examine whether the online LiRA variant with a strict nominal FPR of 0.001\% produces consistent and stable results \emph{in all runs} of CIFAR-10. 
We focus on the online variant because the authors of LiRA~\cite{carlini2022membership} consider it to be most effective when the number of shadow models exceeds 64 (we use 256).
We ran multiple independent training runs (fixed training/test split; see Section~\ref{sec:evaluation_metrics}): 
(i) \emph{identical settings}--12 runs with different seeds but identical hyperparameters; 
(ii) \emph{minor variations}--batch size \(=512\) with dropout \(=0.2\), 
and MixUp augmentation (instead of CutMix) with dropout \(=0.15\); 
(iii) \emph{architectural change}--ResNet18 vs.\ WideResNet28-2; 
(iv) \emph{transfer learning}; 
and (v) \emph{combined variations}.

\begin{figure*}[t!]
  \centering
  \begin{subfigure}[t]{0.33\textwidth}
    \centering
    \includegraphics[width=\linewidth]{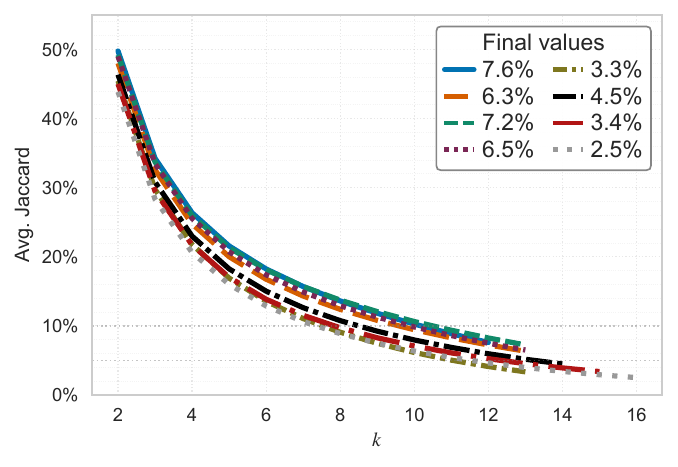}
    \caption{\textbf{Reproducibility vs.\ $k$ (Avg.\ Jaccard)}}
  \end{subfigure}\hfill
  \begin{subfigure}[t]{0.33\textwidth}
    \centering
    \includegraphics[width=\linewidth]{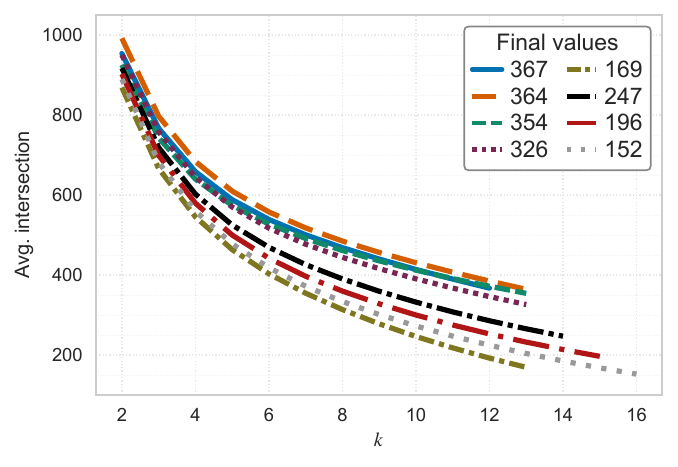}
    \caption{\textbf{Stability (Intersection)}}
  \end{subfigure}\hfill
  \begin{subfigure}[t]{0.33\textwidth}
    \centering
    \includegraphics[width=\linewidth]{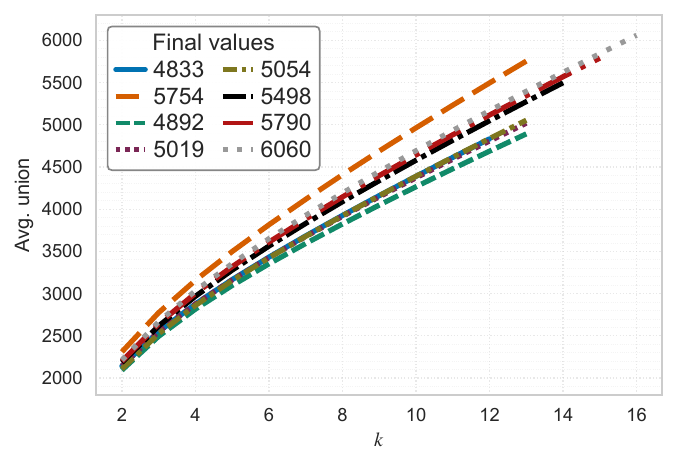}
    \caption{\textbf{Coverage (Union)}}
  \end{subfigure}
  \includegraphics[width=0.65\textwidth]{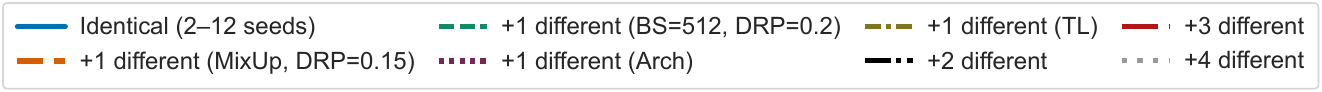}
  \caption{Reproducibility, stability, and coverage vs seeds, training variations, and runs (TP$\geq$1).}
  \Description{Three line plots summarizing reproducibility online LiRA attack results across multiple training runs. 
            Panel (a) shows Jaccard similarity decreases sharply as runs increase. 
            Panel (b) shows intersection shrinks with more runs. 
            Panel (c) shows union expands with more runs and training variations. 
            Together, these plots illustrate that vulnerable sample identification becomes increasingly inconsistent across seeds and configurations.}
    \label{fig:stability_coverage_001}
\end{figure*}

\paragraph{Severe inconsistency across runs} 
Fig.~\ref{fig:stability_coverage_001} shows that, as more runs combined, 
the intersection of vulnerable samples (those identified in all runs) shrank sharply, while the union (samples identified in any run) expanded rapidly. 
This divergence indicates that the identity of thresholded ``vulnerable'' samples is highly sensitive to retraining variability, rather than converging to a small, stable set.
Even for {\em identical settings} with only random seed variation (blue curve), \emph{Jaccard similarity} dropped from about $50\%$ agreement at $k{=}2$ runs, to less than $7.6\%$ at $k{=}12$. 
The decay accelerated when additional variations were introduced, and combining multiple variations yielded near-zero reproducibility --an average Jaccard similarity of $2.5\%$ for four distinct configurations ($k{=}16$). 
In other words, more than $97\%$ of the samples flagged as ``vulnerable'' in any single run were not consistently identified in others. 
This dramatic drop arises from sensitivity at the target FPR boundary: even if a single model assigns an unusually high confidence to a sample (member or non-member), it can push that sample above the decision threshold and include it in the vulnerable set for that run.
This causes near-boundary samples to fluctuate across runs, expanding the union set.
Since the union curves show no sign of convergence, it appears that adding more runs would eventually label a large fraction of training samples as vulnerable.

\begin{figure*}[t!]
    \centering
    \includegraphics[width=\linewidth]{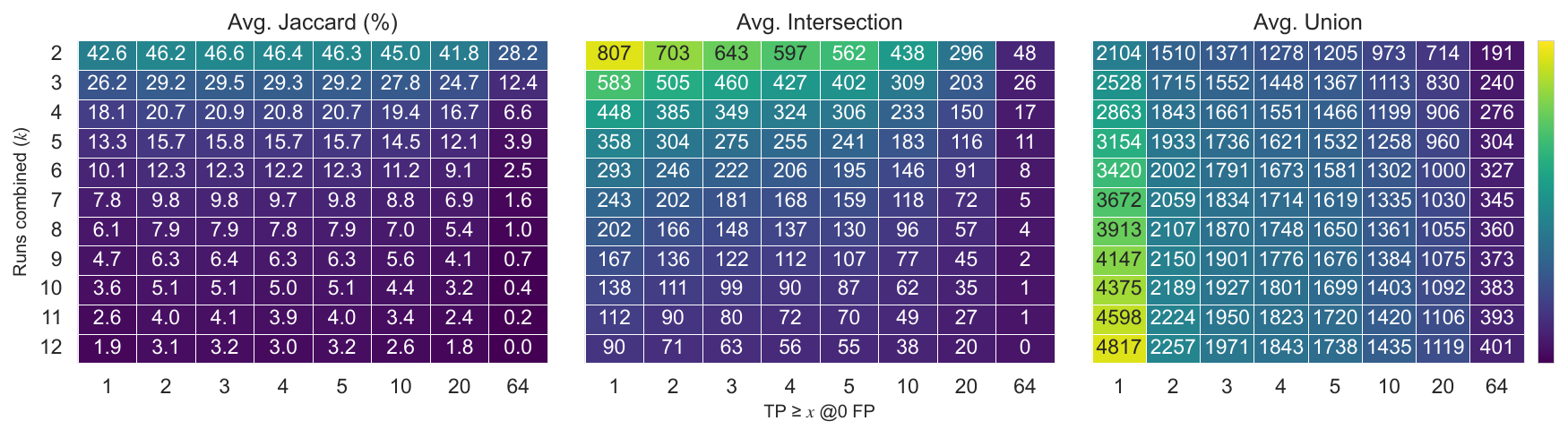}
    \caption{Reproducibility, stability,
    and coverage of zero-FP LiRA positives across runs (identical settings).
    Rows: number of \emph{runs combined} (\(k\)). Columns: within-run support threshold
    \(x\) (TP\(\ge x\) among 128 IN shadows), all at FP\(=0\).
    Modest support (\(x\in[2,5]\)) improves 
    reproducibility, while very strong support (\(x\ge20\)) reduces both stability and coverage.}
    \Description{A grid of heatmaps showing reproducibility, stability, and coverage of zero-false-positive LiRA detections across multiple runs. 
        Each row corresponds to the number of runs combined (k), and each column to the within-run support threshold x, defined as the number of agreeing shadow models among 128 IN models. 
        Moderate support thresholds between 2 and 5 improve reproducibility, while higher thresholds reduce both stability and coverage.}
    \label{fig:tp_geq_identical_001}
\end{figure*}

\paragraph{Support thresholds and the coverage–stability tradeoff}
Fig.~\ref{fig:tp_geq_identical_001} analyzes the \emph{zero-FP} detections while varying the support threshold \(x\), \emph{i.e.}, the number of IN shadow models (out of 128) that must agree within a run.  
We can see that increasing the support threshold yielded only limited gains in reproducibility.  
The average Jaccard similarity rose slightly from \(\,1.9\%\) at \(x{=}1\) to a maximum of \(\approx3.2\%\) for \(x\in[2,5]\), after which reproducibility declined again.  
For any fixed \(x\), agreement decreased sharply as more runs were combined, indicating substantial variability across runs even under FP\(\;=0\).
Increasing the support threshold also substantially reduced stability and coverage.  
At \(k{=}12\), the average intersection shrank from 90 shared positives at \(x{=}1\) to 0 at \(x{=}64\).  
The average union also decreased from 4817 detections at \(x{=}1\) to 401 at \(x{=}64\).
Even with FP\(\;=0\) per run, most low-support detections (TP\(\;=1\)) are run-specific and fail to generalize across seeds; they reflect edge cases driven by noise and uncertainty in shadow-based thresholds. 
Support levels in the range \(x\in[2,5]\) substantially reduced the size of the union with only a small to modest decrease in the intersection due to the elimination of noisy detections.
Therefore, these support levels represent the most informative range (TP\(\ge x\)@0FP) to assess reproducibility.

\begin{figure*}[t!]
    \centering
    \includegraphics[width=\linewidth]{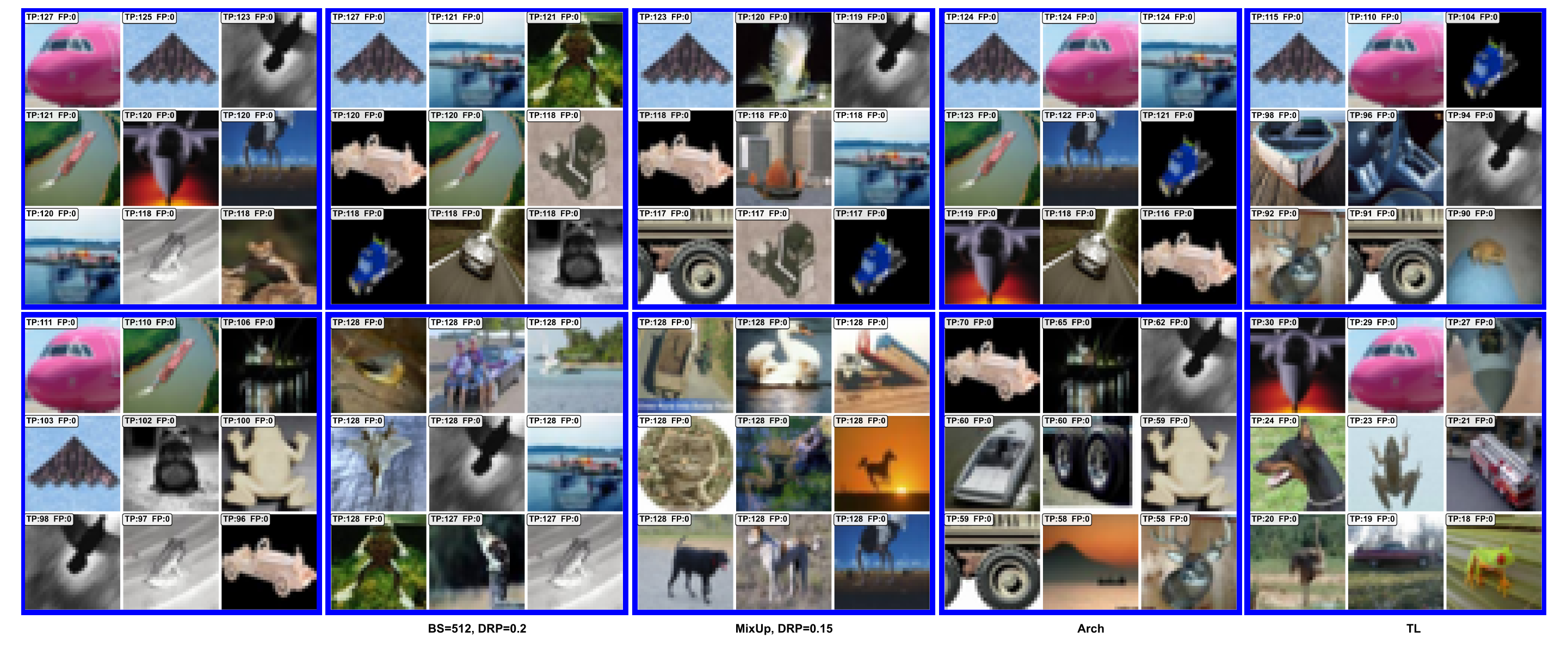}
    \caption{
        Top-9 most vulnerable samples in 10 different runs. Upper row and leftmost run in the lower row: six independent runs 
        (identical settings, different seeds). Rest of runs in the lower row, respectively: one run with a slight change in batch size and dropout, 
        one run using MixUp instead of CutMix together with a modest dropout increase, 
        one run using a different but similar architecture (ResNet18 vs.\ WideResNet28-2), 
        and one run with transfer learning. 
        On each image, TP denotes the number of true detections (true positives)
        among 128 IN shadow models in that run, while FP denotes the number of false
        detections among the 128 OUT shadow models (FP$=0$ for all displayed samples).}
     \Description{
        The figure shows multiple grids of small image patches, each grid presenting the nine samples most frequently flagged as vulnerable by the online LiRA attack in a particular training run. 
        Each grid is arranged in a 3×3 layout. 
        For each image, TP denotes the number of true detections among 128 IN shadow models in that run, and FP denotes the number of false detections among the 128 OUT shadow models (FP=0 for all shown samples).
        The grids in the top row and the leftmost grid of the bottom row come from runs with identical training settings but different random seeds, yet they display noticeably different sets of images with little overlap in sample identity or ranking. 
        The remaining grids on the bottom row correspond to runs with variations in batch size and dropout, a run using MixUp instead of CutMix, a run with a different model architecture, and a run using transfer learning. 
        Overall, the figure illustrates that the samples identified as most vulnerable vary widely across runs and configurations, highlighting the instability of sample-level vulnerability.}

    \label{fig:top_k_001}
\end{figure*}

\paragraph{Instability of top-ranked vulnerable samples}
Even among the 9 most confidently identified vulnerable samples (\emph{i.e.}, those with the highest support within FP$\;=0$ runs, see Fig.~\ref{fig:top_k_001}), there is little agreement between the runs on the identity or ranking of the samples. 
New samples frequently appeared among the top-ranked vulnerabilities despite identical training settings. With more setting variations, disagreement increased more.
The results in Fig.~\ref{fig:tp_geq_identical_001} and Fig.~\ref{fig:top_k_001} indicate that \textbf{``vulnerability'' should \emph{not} be viewed as an intrinsic property of specific samples, but is also influenced by other factors, including model generalization and stochastic training factors such as initialization, mini-batch ordering, and local neighborhood effects~\cite{ye2022enhanced,wang2025membership}.} 
These factors can reshuffle samples across runs, especially those near the extreme low-FPR boundary.

In general, these reproducibility results show that 
\textbf{LiRA’s \emph{thresholded} sets at extremely low FPR exhibit limited reproducibility across runs, reducing their reliability for identifying a small, stable set of ``vulnerable'' samples from a single run.}

\begin{figure}[t!]
\centering
\includegraphics[width=\linewidth]{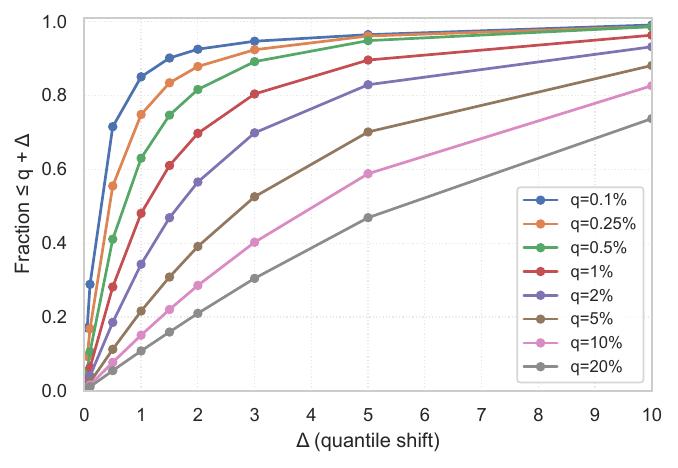}
\caption{Rank displacement under run-to-run variability.
  Curves show the fraction of displaced samples whose percentile rank remained within the Top-$(q{+}\Delta)\%$.}
    \Description{The figure shows a set of curves of the fraction of displaced samples that remain near the Top-$q\%$ cutoff as the tolerance $\Delta$ increases. The horizontal axis is $\Delta$ (percentage points added to the cutoff), and the vertical axis is the fraction of displaced samples whose percentile rank in the reference run remains within the Top-$(q+\Delta)\%$. Each curve corresponds to a different Top-$q\%$ level (as indicated by the legend). Error bars summarize variability across runs (mean $\pm$ standard deviation). The curves rise quickly with small $\Delta$, indicating that most displaced samples fall just below the cutoff rather than far down the ranking.}
\label{fig:rank_displacement}
\end{figure}

\paragraph{Granular score-level validity under run variability}
Beyond thresholded sets at a target FPR, we examine whether LiRA’s likelihood ratios themselves provide a stable vulnerability ranking signal.
For each run, we compute a per-sample score using the \texttt{median gap} (Section~\ref{sec:evaluation_metrics}).
We favor the median over the mean since the \texttt{mean gap} exhibits higher variance and weaker extreme-tail stability (see Appendix~\ref{app:more_repreduce}).
Table~\ref{tab:fpr_vs_topq_median_gap} reports reproducibility of ranking-based Top-$q\%$ sets induced by the \texttt{median gap} under two notions of stability:
\emph{pairwise} agreement averaged over all pairs of runs ($k{=}2$), and \emph{all-runs} agreement requiring membership in all $K{=}12$ runs ($k{=}12$).
We also compare these sets to the corresponding FPR-thresholded selections.
The global Spearman correlation over all samples was 83.5\%$_{\pm 5.0\%}$, indicating that the overall vulnerability ranking was largely preserved across runs. 
\begin{table}[t!]
\caption{Reproducibility of ranking-based sets (Top-$q\%$ by \texttt{median gap}) and FPR-thresholded sets.
$\rho$ denotes the Spearman rank correlation, computed on the intersection set.
$^\dagger$/$^\ddagger$: Top-$q\%$ closely matching the all-runs $|\cap|$ of FPR\,$=\!0.001\%$ / $0.1\%$.}
\label{tab:fpr_vs_topq_median_gap}
\centering
\small
\resizebox{\linewidth}{!}{
\begin{tabular}{llcccccc}
\toprule
& & \multicolumn{3}{c}{Pairwise ($k=2$ runs)} & \multicolumn{3}{c}{All-runs ($k=12$ runs)} \\
\cmidrule(lr){3-5}\cmidrule(lr){6-8}
Top-$q\%$ & $T_q$
& Tail $\rho$ (\%) & $|\cap|/|\cup|$ & Jaccard (\%)
& Tail $\rho$ (\%) & $|\cap|/|\cup|$ & Jaccard (\%) \\
\midrule
0.10  & 50    & 49.45$_{\pm 17.95}$ & 26/74$_{\pm 7}$         & 36.02$_{\pm 13.33}$ & 49.44 & 6/158         & 3.80  \\
0.25  & 125   & 46.25$_{\pm 15.68}$ & 72/178$_{\pm 16}$       & 41.25$_{\pm 12.88}$ & 55.85 & 21/338        & 6.21  \\
0.50  & 250   & 47.74$_{\pm 16.24}$ & 157/343$_{\pm 26}$      & 46.63$_{\pm 11.29}$ & 46.39 & 56/621        & 9.02  \\
0.75  & 375   & 50.58$_{\pm 15.30}$ & 242/508$_{\pm 38}$      & 48.58$_{\pm 11.23}$ & 46.77 & 92/892        & 10.31 \\
1.00  & 500   & 53.45$_{\pm 13.79}$ & 330/670$_{\pm 51}$      & 50.16$_{\pm 11.27}$ & 51.67 & 134/1153      & 11.62 \\
2.00  & 1000  & 58.93$_{\pm 12.44}$ & 721/1279$_{\pm 92}$     & 57.11$_{\pm 10.73}$ & 57.01 & 327/2050      & 15.95 \\
2.24$^\dagger$ & 1121 & 59.70$_{\pm 12.56}$ & 816/1426$_{\pm 102}$  & \textbf{57.96}$_{\pm 10.63}$ & 57.87 & \textcolor{red}{368}/\textbf{2280} & \textbf{16.14} \\
3.00  & 1500  & 62.16$_{\pm 12.43}$ & 1123/1877$_{\pm 126}$   & 60.46$_{\pm 10.06}$ & 62.43 & 540/2887      & 18.70 \\
4.00  & 2000  & 64.54$_{\pm 12.00}$ & 1539/2461$_{\pm 161}$   & 63.20$_{\pm 9.99}$  & 63.71 & 807/3675      & 21.96 \\
5.00  & 2500  & 66.88$_{\pm 11.99}$ & 1970/3030$_{\pm 193}$   & 65.63$_{\pm 9.90}$  & 65.88 & 1103/4392     & 25.11 \\
10.00 & 5000  & 74.62$_{\pm 11.61}$ & 4157/5843$_{\pm 320}$   & 71.61$_{\pm 8.82}$  & 73.17 & 2633/7851     & 33.54 \\
10.38$^\ddagger$ & 5190 & 74.97$_{\pm 11.58}$ & 4327/6053$_{\pm 331}$ & \textbf{71.95}$_{\pm 8.84}$ & 73.36 & \textcolor{blue}{2749}/\textbf{8101} & \textbf{33.93} \\
20.00 & 10000 & 81.20$_{\pm 10.63}$ & 8745/11255$_{\pm 460}$  & 77.98$_{\pm 7.01}$  & 80.71 & 6426/13993    & 45.92 \\
30.00 & 15000 & 84.99$_{\pm 9.19}$  & 13302/16698$_{\pm 503}$ & 79.82$_{\pm 5.33}$  & 84.20 & 10367/20344   & 50.96 \\
40.00 & 20000 & 87.39$_{\pm 7.36}$  & 17705/22295$_{\pm 573}$ & 79.53$_{\pm 4.60}$  & 86.62 & 13958/27650   & 50.48 \\
50.00 & 25000 & 88.80$_{\pm 5.93}$  & 21805/28195$_{\pm 884}$ & 77.51$_{\pm 5.59}$  & 88.25 & 16963/37088   & 45.74 \\
60.00 & 30000 & 89.66$_{\pm 4.94}$  & 25557/34443$_{\pm 1092}$& 74.37$_{\pm 5.55}$  & 89.18 & 19241/45857   & 41.96 \\
70.00 & 35000 & 90.10$_{\pm 4.18}$  & 29411/40589$_{\pm 882}$ & 72.54$_{\pm 3.74}$  & 89.71 & 21140/49653   & 42.58 \\
80.00 & 40000 & 89.79$_{\pm 3.74}$  & 34382/45618$_{\pm 525}$ & 75.39$_{\pm 2.02}$ & 90.09  & 23272/49998   & 46.55 \\
90.00 & 45000 & 87.75$_{\pm 3.97}$  & 41147/48853$_{\pm 193}$ & 84.23$_{\pm 0.73}$  & 90.65 & 27310/50000   & 54.62 \\
100.00& 50000 & 83.46$_{\pm 5.00}$  & 50000/50000$_{\pm 0}$   & 100.00$_{\pm 0.00}$ & 83.46 & 50000/50000   & 100.00 \\
\midrule
FPR $=0.001\%$ & --- & --- & 954/2142   & 49.80 & --- & \textcolor{red}{367}/4833   & 7.60  \\
FPR $=0.1\%$   & --- & --- & 4239/8454  & 52.10 & --- & \textcolor{blue}{2748}/18132 & 15.20 \\
\bottomrule
\end{tabular}}
\end{table}
For small $q$, pairwise agreement was moderate to high, but agreement across all 12 runs was much stricter. 
For example, at $q{=}1\%$ ($|T_q|{=}500$), the average pairwise intersection was $330\pm 51$ (Jaccard $50.16\%\pm 11.27\%$), while the all-runs intersection contained 134 samples (Jaccard $11.62\%$).
At $q{=}0.1\%$, where selections were extremely small, all-runs agreement was much stricter (pairwise Jaccard $36.02\%\pm 13.33\%$ vs.\ all-runs $3.80\%$).
As $q$ increased, both notions improved (\emph{e.g.}, $q{=}10\%$: pairwise Jaccard $71.61\%\pm 8.82\%$ and all-runs $33.54\%$), consistently with broader high-risk regions being more reproducible.
Hence, strict identification in the extreme Top-$q\%$ remains the most sensitive regime, and thus a single-run selection may omit samples that rank highly in another run.
Disagreement \emph{accumulates} when aggregating across many runs:
at $q{=}1\%$, the average pairwise union was 670 (a $+34\%$ expansion over $|T_q|$), while the union over all $K{=}12$ runs expanded to 1153 (a $+131\%$ expansion), reflecting boundary disagreements that are modest for a pair of run but compound across runs.
Figure~\ref{fig:rank_displacement} quantifies how far \emph{displaced} samples fall below the Top-$q\%$ cutoff, distinguishing large rank shifts from near-cutoff churn.
For each run $r$, we consider samples that appear in the Top-$q\%$ set of at least one run but not in run $r$, and measure their percentile rank in run $r$.
We then report, for each $\Delta$, the fraction of displaced samples whose percentile rank remains within $q+\Delta$ percentage points.
Displacement was predominantly local (especially in the extreme tail): most displaced samples fell just below the cutoff rather than far down the ranking.
For instance, at $q{=}1\%$, $69.73\%\pm 4.85\%$ of displaced samples remained within $q{+}2\%$ and $89.57\%\pm 1.98\%$ within $q{+}5\%$.
At the more extreme $q{=}0.1\%$, $92.52\%\pm 2.69\%$ remained within $q{+}2\%$.
This locality helps explaining why unions expand under all-runs aggregation even when cross-run rank changes are mostly near the cutoff.
In practice, stability can be improved by selecting a larger Top-$q\%$ set in a single run (broader but more reproducible) or by aggregating across runs (\emph{e.g.}, taking the union), at additional computational cost.
Figure~\ref{fig:median_gap_bins} in Appendix~\ref{app:more_repreduce} provides complementary evidence that score dispersion was the largest in the extreme tail.

Matching methods by the strictly stable core (all-runs $|\cap|$, $k{=}12$) highlights a clear practical advantage of ranking-based selection.
For comparable stable cores, ranking yielded substantially smaller unions and higher all-runs Jaccard than FPR-thresholding:
at FPR $=0.001\%$, thresholding yielded $367/4833$ (Jaccard $7.6\%$), whereas the matched ranking-based choice ($q{=}2.24\%$) yielded $368/2280$ (Jaccard $16.14\%$);
at FPR $=0.1\%$, thresholding yielded $2748/18132$ (Jaccard $15.2\%$), whereas the matched ranking-based choice ($q{=}10.38\%$) yielded $2749/8101$ (Jaccard $33.93\%$).
Thus, for a fixed stable core size relevant to targeted mitigation, ranking-based selection produced more stable and consistent cross-run sets (smaller union expansion and higher all-runs Jaccard).

Overall, \textbf{the likelihood-ratio scores produced by online LiRA retain meaningful rankings that are more stable and consistent across runs than the FPR-thresholded sets, suggesting that LiRA is more reliable as a ranking-based auditing tool than as a precise single-run selector at extreme FPR thresholds}.

\subsection{Results with Target FPR of 0.1\%}
\label{sec:realistic-0p1}

\begin{figure*}[t!]
  \centering
  \begin{subfigure}[t]{0.33\textwidth}
    \centering
    \includegraphics[width=\linewidth]{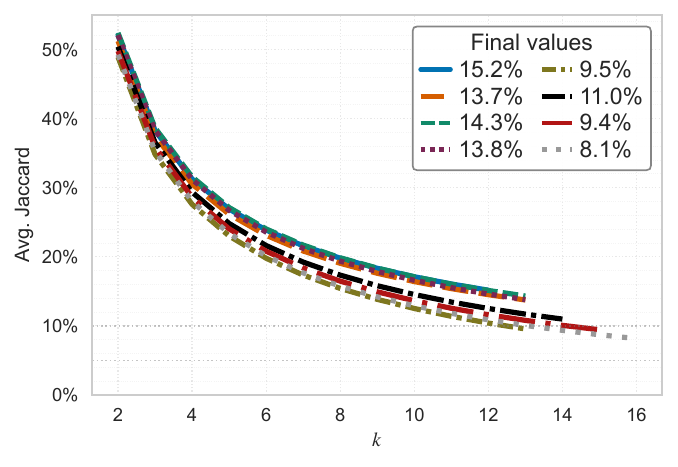}
    \caption{\textbf{Reproducibility vs.\ $k$ (Avg.\ Jaccard)}}
  \end{subfigure}\hfill
  \begin{subfigure}[t]{0.33\textwidth}
    \centering
    \includegraphics[width=\linewidth]{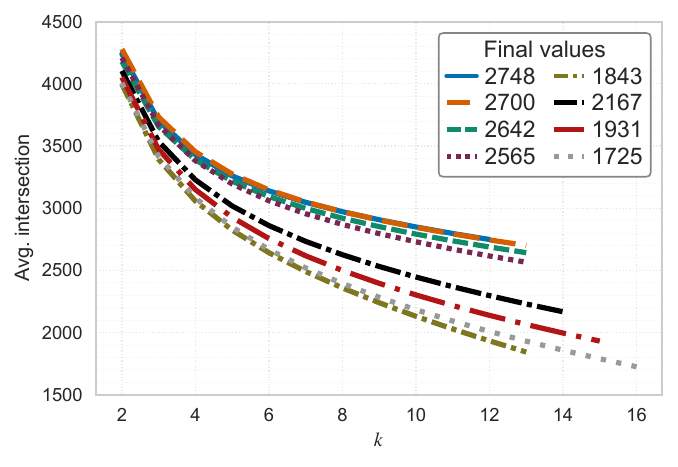}
    \caption{\textbf{Stability (Intersection)}}
  \end{subfigure}\hfill
  \begin{subfigure}[t]{0.33\textwidth}
    \centering
    \includegraphics[width=\linewidth]{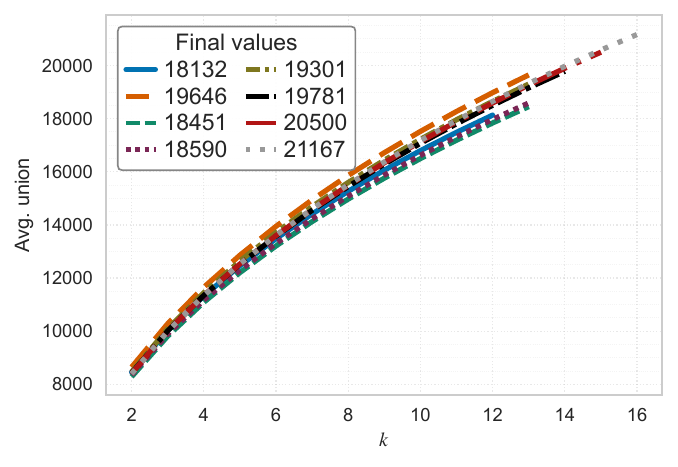}
    \caption{\textbf{Coverage (Union)}}
  \end{subfigure}
  \caption{Reproducibility, stability, and coverage vs seeds, training variations, and runs (TP$\geq$1 for 0.1\%FPR).}
  \Description{Three line plots summarizing reproducibility online LiRA attack results across multiple training runs for 0.1\%FPR. 
            Panel (a) shows Jaccard similarity decreases sharply as runs increase. 
            Panel (b) shows intersection shrinks with more runs. 
            Panel (c) shows union expands with more runs and training variations. 
            Together, these plots illustrate that vulnerable sample identification becomes increasingly inconsistent across seeds and configurations.}
    \label{fig:stability_coverage_01}
\end{figure*}

Table~\ref{tab:cifar10_both_0p1} shows the results for CIFAR-10 at a target FPR of $0.1\%$ under both optimistic (target-based) and realistic (shadow-based) calibration. 
As expected, the relaxation of the nominal FPR significantly degraded the PPV for all attack variants for these settings. 
Fig.~\ref{fig:stability_coverage_01} presents the corresponding reproducibility analysis. 
While the more tolerant FPR yielded higher apparent stability than the 0.001\% setting, it also substantially increased the size of the union set. 
Many samples were flagged as vulnerable in at least one run, yet only a small subset was consistently identified across runs.
Thus, increasing the target FPR improves agreement at the cost of substantially broader vulnerable sets, reducing selectivity for targeted mitigation. 
Additional analysis of zero-FP detections under varying within-run support thresholds is presented in Appendix~\ref{app:more_repreduce} (Fig.~\ref{fig:tp_geq_identical_01}).
\begin{table}[t!]
\caption{CIFAR-10 under target vs.\ shadow calibration (target FPR $=0.1\%$).}
\label{tab:cifar10_both_0p1}
\centering
\small
\resizebox{\linewidth}{!}{
\begin{tabular}{ll|cc|ccc}
\toprule
 & & \multicolumn{2}{c|}{Performance} & \multicolumn{3}{c}{PPV (\%)} \\
Benchmark & Attack & TPR$'$ (\%)& FPR$'$ (\%)& @$\pi$=1\% & @$\pi$=10\% & @$\pi$=50\% \\
\midrule
\multicolumn{7}{l}{\textit{\textbf{Target-based thresholds (optimistic)}}} \\
\cmidrule(lr){1-7}
\multirow{4}{*}{Baseline}
& Online       & 10.268$_{\pm 0.555}$ & 0.098$_{\pm 0.001}$ & 51.271$_{\pm 1.378}$ & 92.038$_{\pm 0.408}$ & 99.048$_{\pm 0.053}$ \\
& Online (FV)  &  9.135$_{\pm 0.508}$ & 0.098$_{\pm 0.001}$ & 48.350$_{\pm 1.414}$ & 91.137$_{\pm 0.461}$ & 98.931$_{\pm 0.060}$ \\
& Offline      &  3.262$_{\pm 0.338}$ & 0.098$_{\pm 0.001}$ & 25.031$_{\pm 1.963}$ & 78.499$_{\pm 1.803}$ & 97.040$_{\pm 0.310}$ \\
& Offline (FV) &  4.540$_{\pm 0.424}$ & 0.098$_{\pm 0.001}$ & 31.724$_{\pm 2.028}$ & 83.573$_{\pm 1.313}$ & 97.860$_{\pm 0.202}$ \\
\cmidrule(lr){1-7}
\multicolumn{7}{l}{\textit{\textbf{Shadow-based thresholds (realistic)}}} \\
\cmidrule(lr){1-7}
\multirow{4}{*}{Baseline}
& Online       & 10.245$_{\pm 0.372}$ & 0.101$_{\pm 0.022}$ & 51.227$_{\pm 5.294}$ & 91.899$_{\pm 1.573}$ & 99.027$_{\pm 0.204}$ \\
& Online (FV)  &  9.151$_{\pm 0.347}$ & 0.101$_{\pm 0.022}$ & 48.444$_{\pm 5.079}$ & 91.036$_{\pm 1.657}$ & 98.914$_{\pm 0.219}$ \\
& Offline      &  3.261$_{\pm 0.165}$ & 0.102$_{\pm 0.025}$ & 25.271$_{\pm 4.670}$ & 78.267$_{\pm 4.208}$ & 96.973$_{\pm 0.738}$ \\
& Offline (FV) &  4.520$_{\pm 0.213}$ & 0.100$_{\pm 0.022}$ & 32.195$_{\pm 5.124}$ & 83.561$_{\pm 3.106}$ & 97.845$_{\pm 0.477}$ \\
\cmidrule(lr){2-7}
\multirow{4}{*}{AOF}
& Online       &  2.826$_{\pm 0.750}$ & 0.189$_{\pm 0.871}$ & 21.571$_{\pm 7.179}$ & 72.813$_{\pm 11.409}$ & 95.448$_{\pm 4.870}$ \\
& Online (FV)  &  3.489$_{\pm 0.404}$ & 0.103$_{\pm 0.032}$ & 26.554$_{\pm 5.090}$ & 79.176$_{\pm 6.060}$ & 97.007$_{\pm 2.649}$ \\
& Offline      &  0.855$_{\pm 5.037}$ & 0.434$_{\pm 5.077}$ &  5.349$_{\pm 1.715}$ & 37.481$_{\pm 7.781}$ & 83.563$_{\pm 5.673}$ \\
& Offline (FV) &  1.665$_{\pm 3.026}$ & 0.296$_{\pm 3.086}$ & 13.403$_{\pm 2.987}$ & 62.180$_{\pm 7.044}$ & 93.274$_{\pm 4.969}$ \\
\cmidrule(lr){2-7}
\multirow{4}{*}{AOF+TL}
& Online       &  0.559$_{\pm 0.086}$ & 0.112$_{\pm 0.056}$ &  5.244$_{\pm 1.278}$ & 37.323$_{\pm 6.238}$ & 83.761$_{\pm 4.237}$ \\
& Online (FV)  &  0.826$_{\pm 0.102}$ & 0.101$_{\pm 0.025}$ &  7.946$_{\pm 1.673}$ & 48.212$_{\pm 5.353}$ & 89.161$_{\pm 2.022}$ \\
& Offline      &  0.126$_{\pm 0.128}$ & 0.135$_{\pm 0.132}$ &  0.973$_{\pm 0.341}$ &  9.669$_{\pm 2.850}$ & 48.176$_{\pm 7.251}$ \\
& Offline (FV) &  0.305$_{\pm 0.119}$ & 0.133$_{\pm 0.115}$ &  2.684$_{\pm 0.861}$ & 22.890$_{\pm 5.747}$ & 71.679$_{\pm 6.875}$ \\
\bottomrule
\end{tabular}}
\end{table}

\subsection{Loss Ratio as a Predictor}
\label{sec:lratio}
\begin{figure}[t!]
    \centering
    \includegraphics[width=0.85\linewidth]{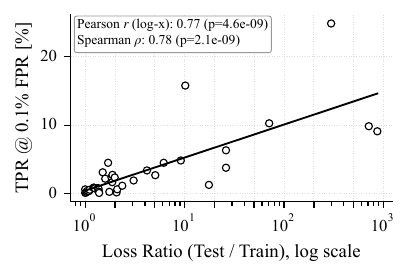}
    \caption{Correlation between loss ratio (test loss / train loss, log scale) and online LiRA TPR@0.1\% FPR. Points represent 41 different model configurations}
         \Description{The figure is a scatter plot where each point corresponds to one of 41 model configurations evaluated under different datasets, architectures, and regularization settings. 
             The horizontal axis shows the loss ratio on a logarithmic scale, and the vertical axis shows the online LiRA TPR@0.1\% FPR. 
             The points form an upward trend, indicating that configurations with higher loss ratios also tend to yield higher LiRA success. 
             Models with higher loss ratios tend to exhibit higher attack success rates, while models with ratios below 2 are much less vulnerable.}
    \label{fig:lossratio_corr}
\end{figure}

Across 41 configurations, we observed a clear monotonic relationship between a model’s vulnerability to online LiRA and its \emph{loss ratio} (see Figure~\ref{fig:lossratio_corr}). 
Models with larger loss ratios consistently exhibited higher LiRA success rates, while well-generalized models with ratios below 2 were much less vulnerable. 
This trend holds across datasets, architectures, and regularization settings, suggesting that the loss ratio could serve as a simple, task-agnostic proxy for privacy risk. 
It captures in a single quantity how far a model’s behavior departs from perfect generalization, linking overfitting directly to the strength of the membership signals available to the attacker.

\section{Discussion}
\label{sec:discussion}

\paragraph{Overconfidence dominates.}
Models with high prediction certainty on training data ({\em i.e.}, loss-wise overfitted) are more vulnerable to attack, 
regardless of whether shadow-based threshold or skewed priors are applied. 
Their per-sample losses are often polarized and similar across runs. 
This observation aligns with the empirical findings in~\cite{overconfidence2024}, 
who show that the separation between members and non-members increases with the prediction confidence of the model on members. 
However, evaluating LiRA on such overconfident models deviates from realistic deployment and substantially exaggerates its actual privacy threat.

\begin{figure}[t!]
    \centering
    \begin{subfigure}[t]{\linewidth}
        \centering
        \includegraphics[width=0.9\linewidth]{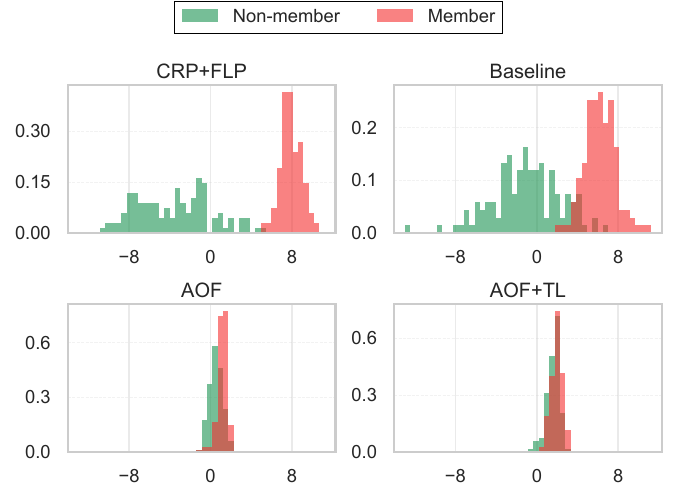}
        \caption{Scaled logit scores}
        \label{fig:sample_inout_score}
    \end{subfigure}%
    
    \begin{subfigure}[t]{\linewidth}
        \centering
        \includegraphics[width=0.9\linewidth]{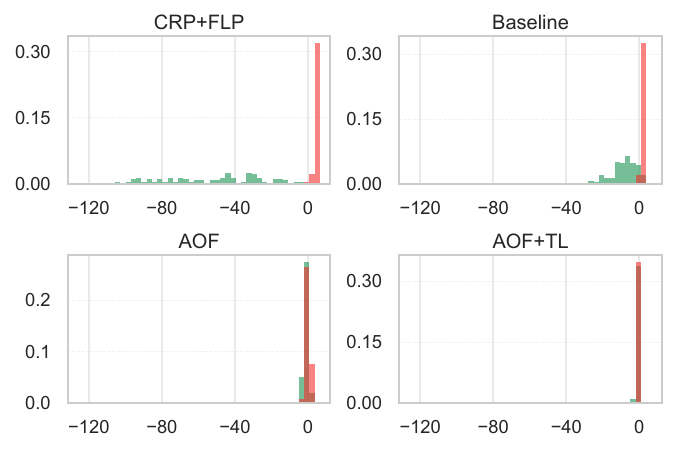}
        \caption{Log-likelihood ratios}
        \label{fig:sample_inout_online_ratio}
    \end{subfigure}
    \caption{In/out distributions for a representative CIFAR-10 sample under the online attack. 
            AOF and TL significantly narrow the gap between members and non-members.}
         \Description{The figure shows two sets of distributions for a single CIFAR-10 sample comparing member and non-member scores under different training configurations. 
                 In panel (a), the scaled logit score distributions for IN and OUT samples overlap increasingly as AOF and transfer learning are applied, reducing their separation. 
                 Panel (b) shows the corresponding log-likelihood ratios, which also collapse into a tightly overlapping range. 
                 Together, the plots illustrate that stronger regularization makes IN and OUT signals nearly indistinguishable for this sample.}
    \label{fig:sample_inout_dist}
\end{figure}

\paragraph{Substantial degradation under realistic conditions}
When evaluated under realistic conditions (models trained with AOF or TL to reduce prediction gaps, thresholds calibrated from shadow models, and skewed priors, that is, $\pi \leq 10\%$), 
the apparent success of LiRA substantially degrades.
Membership predictions become less reliable and less reproducible because AOF and TL compress the confidence distributions of members and non-members, which LiRA models as Gaussians.
As these distributions get closer to each other, the likelihood that the signal of a target model originates from an \emph{OUT} sample increases, pushing thresholds to extreme values and reducing TPR while inflating the achieved FPR$'$ variance.  
Fig.~\ref{fig:sample_inout_dist} illustrates this effect for a representative CIFAR-10 sample.
AOF and TL progressively narrow the gap between scaled logit scores, which prevents LiRA from effectively distinguishing between the sample likelihood ratios. 
Inference-time augmentation can strengthen the membership signal~\cite{carlini2022membership}, but its effect fades away when models are not overfitted. 
Consequently, thresholds estimated from non-overfitted shadow models transfer poorly, producing large gaps between nominal and achieved FPRs. 
This calibration-induced variance propagates through the Bayes rule into PPV, making per-sample inferences less reliable even when mean TPRs remain moderate. 
This is not a flaw of LiRA but a statistical inevitability: once member and non-member confidence distributions overlap substantially, 
no single-sample decision rule can achieve both low FPR and high precision~\cite{zhu2025impact}.

\paragraph{Limited reproducibility.} 
Across 12 runs with identical configurations but different seeds, the average Jaccard similarity between thresholded sets was $7.6\%$ at 0.001\% FPR and $15.2\%$ at 0.1\% FPR, while the union of flagged samples steadily expanded (especially at 0.1\% FPR).   
This indicates that the \emph{identity of thresholded samples at extreme FPR} is highly sensitive to run-to-run variability, even when aggregate accuracy and loss remain nearly unchanged. 
In contrast, the underlying LiRA likelihood ratios retain a meaningful and relatively stable \emph{ vulnerability ranking} signal.
While strict agreement at small Top-$q\%$ selections remains limited (especially under all-runs intersection), most disagreement is concentrated near the cutoff rather than driven by large rank reversals (Figure~\ref{fig:rank_displacement}).
Practically, these findings suggest that LiRA is better interpreted as a ranking-based auditing tool under retraining variability rather than as a precise single-run selector at extreme thresholds.
Stability can be improved by widening the selection (larger Top-$q\%$), increasing within-run support requirements, or aggregating across run-to-run, but each option trades off selectivity and/or computational cost.
Importantly, this instability reflects variability \emph{across runs} and does not imply the absence of risk for a fixed deployed model; rather, it limits the reliability of single-run sample-level conclusions.

\paragraph{The deployment paradox and privacy–utility synergy.}
MIAs may constitute genuine privacy threats in domains where membership is sensitive. 
However, these domains demand high accuracy and strong regularization, which weaken LiRA. 
This creates a paradox: \emph{the models most vulnerable to MIAs are those least suitable for deployment in privacy-critical settings.} 
Our GTSRB results illustrate this.
Models that meet deployment standards exhibit natural robustness under realistic thresholds and priors, whereas overfitted baselines exaggerate privacy risk by more than an order of magnitude.  

\paragraph{When LiRA may remain relevant and defenses may fail.}
Our findings do not imply that LiRA is obsolete. 
Under favorable conditions for attackers (severe overfitting, poor calibration, or unrealistic evaluation setups), LiRA may remain effective. 
It is a useful \emph{auditing tool} to upper-bound empirical membership leakage. 
Moreover, practical defenses, such as AOF, TL, and output calibration~\cite{jia2019memguard}, may not fully mitigate attacks when assumptions of data sufficiency and distributional alignment do not hold. 
Specifically: (i)~data scarcity or class imbalance can impair generalization and reintroduce exploitable loss gaps; 
(ii)~domain/temporal shifts between training and non-training data may give attackers an advantage; 
and
(iii)~white-box or semi-white-box access ({\em e.g.}, gradients, internal activations) can amplify leakage;
In such cases, empirical auditing remains essential, and combining LiRA-style probing with formal privacy guarantees offers the strongest protection.
Differential privacy~\cite{dwork2006differential} remains valuable for formal guarantees, but may become unnecessary when models generalize well enough to suppress empirical leakage.

\paragraph{Need for efficient, reliable and consistent MIAs.}
\cite{wang2025membership} proposed ensembling several instances of an attack (or multiple attacks) across runs to improve stability, 
but such approaches are computationally prohibitive for large-scale auditing.
LiRA already incurs substantial computational overhead.
For example, with the equipment mentioned in 
Appendix~\ref{app:additional_setting}, training 256 shadow models on CIFAR-10 with ResNet-18 
took us approximately 29 hours, representing 256× the cost of training a single target model ($\approx 7$ minutes). 
For CIFAR-100 with WideResNet, this increased to 33 hours ($\approx 8$ minutes per model).
Beyond shadow training, LiRA incurred (15-17\%) additional overhead for inference and evaluation when 18 inference augmentations were used. 
This makes the deployment of shadow-heavy attacks on large-scale models computationally less practical, increasing the cost for both auditors and potential attackers.
\cite{wang2025membership} also observed that simple loss-thresholding attacks (the cheapest form of MIA) are often the most consistent across runs, 
even compared to LiRA. 
However, the use of a global loss threshold produces a substantial proportion of false positives~\cite{carlini2022membership}, making such attacks unreliable for per-sample inference. 

\paragraph{Recommendations for defenders and evaluators.}
We recommend:
(i)~training target models with AOF and/or TL to reduce overconfidence and loss–wise overfitting;
(ii)~reporting both the loss ratio and training–evaluation loss curves alongside accuracy to expose leakage risk;
(iii)~evaluating attacks under realistic conditions (shadow-calibrated thresholds and skewed priors, $\pi \leq 10\%$) to reflect an external attacker’s viewpoint; 
and
(iv)~assessing reproducibility across multiple runs to verify the stability of inferred memberships.

\paragraph{Limitations.}
Our study employs small- to moderate-scale discriminative models.
A similar study on large-scale generative and multimodal systems would require many more computational resources and data. 
These larger settings, which have shown limited LiRA performance in prior work~\cite{carlini2022membership}, 
may exhibit different privacy and generalization behaviors. 
We also did not consider other realistic conditions, such as significant data distribution shifts or higher training variance, 
which could further influence attack performance. 
Moreover, we did not evaluate other MIAs beyond LiRA owing to space limitations, 
although our evaluation protocol and conclusions are expected to generalize to similar black-box attacks. 
Finally, loss-ratio values should be interpreted alongside absolute test loss and accuracy, as very low ratios may indicate underfitting rather than genuine robustness.

\paragraph{Ethical principles.} 
This work did not require the review of an external ethics panel.
We evaluated LiRA on publicly available and non-sensitive datasets for realistic privacy assessment, without interacting with human subjects or disclosing sensitive information.

\section{Conclusions and Future Work}
\label{sec:conclusion}
We reviewed LiRA under realistic conditions and found that its effectiveness for membership inference has been overstated. 
When models are properly regularized through anti-overfitting or transfer learning, 
the apparent advantage of LiRA largely decreases without loss of model utility.
Shadow-based calibration and realistic, skewed membership priors reveal that LiRA’s precision (PPV) drops from near-perfect to substantially lower levels for well-generalized models. 
Furthermore, we found that \emph{thresholded} vulnerable sets at extreme low-FPR exhibit limited reproducibility across runs.  
In contrast, LiRA’s likelihood-ratio scores retain a meaningful ranking signal, with instability concentrated in the extreme tail.  
Thus, LiRA is better viewed as a ranking-based auditing tool than as a single-run selector of a small stable subset.  

For practitioners, these findings show that proper use of standard anti-overfitting and/or transfer learning techniques can provide strong empirical privacy protection at no accuracy cost. 
For researchers and evaluators, our results emphasize the need for overconfidence awareness, shadow-based thresholds, realistic priors, and reproducibility checks when quantifying membership risk.

Future work will include applying our evaluation protocol to larger and more diverse data regimes, 
including cross-domain and multimodal settings, 
and exploring adaptive attacks that remain reliable and consistent under realistic evaluation conditions.

\begin{acks}
Partial support to this work has been received from the Government of Catalonia (ICREA Acad\`emia Prizes to J. Domingo-Ferrer and to D. S\'anchez), MCIN/AEI under grant PID2024-157271NB-I00 ``CLEARING-IT'', and the European Commission under project HORIZON-101292277 ``SoBigData IP''.
We used Claude Sonnet 4.5 to optimize code implementation and WriteFull and ChatGPT-5 to correct typos, grammatical errors, and awkward phrasing throughout the article.
\end{acks}

\bibliographystyle{ACM-Reference-Format}
\bibliography{main}

\appendix

\section{Additional Results on Reproducibility}
\label{app:more_repreduce}

This section reports additional reproducibility results.

\paragraph{Reproducibility without zero-FP filtering}
We focus on \emph{all} online LiRA detections on CIFAR-10 (AOF) (\emph{i.e.}, without zero-FP filtering) for both target FPRs.
The patterns align with the results reported in the main paper: reproducibility drops rapidly as more runs are combined, and medium support thresholds improve stabilization.

\begin{figure*}[t!]
    \centering
    \includegraphics[width=\linewidth]{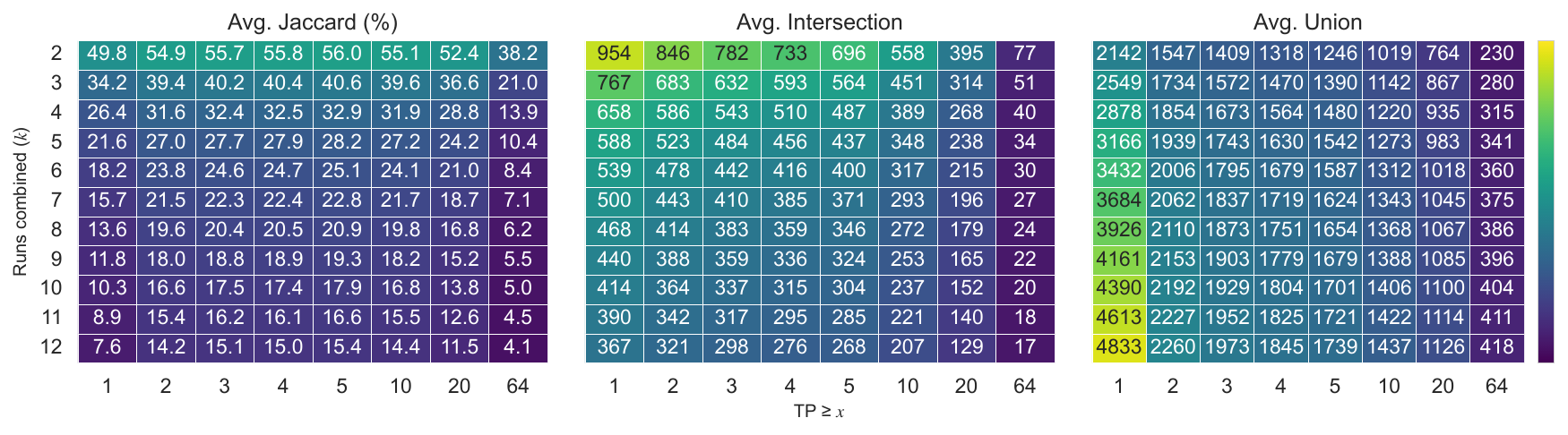}
    \caption{Reproducibility, stability, and coverage of \emph{all} LiRA positives across runs (identical settings) at a target FPR of 0.001\%.}
    \Description{Heatmaps summarizing reproducibility, stability, and coverage for all online LiRA detections at 0.001\% FPR across multiple identical-setting runs. 
                Reproducibility decreases quickly as more runs are combined, and higher support thresholds reduce coverage sharply.}
    \label{fig:tp_geq_all_identical_0001}
\end{figure*}

Fig.~\ref{fig:tp_geq_all_identical_0001} shows results at a target FPR of 0.001\%, where agreement fell rapidly and coverage shrank with increasing support thresholds.

\begin{figure*}[t!]
    \centering
    \includegraphics[width=\linewidth]{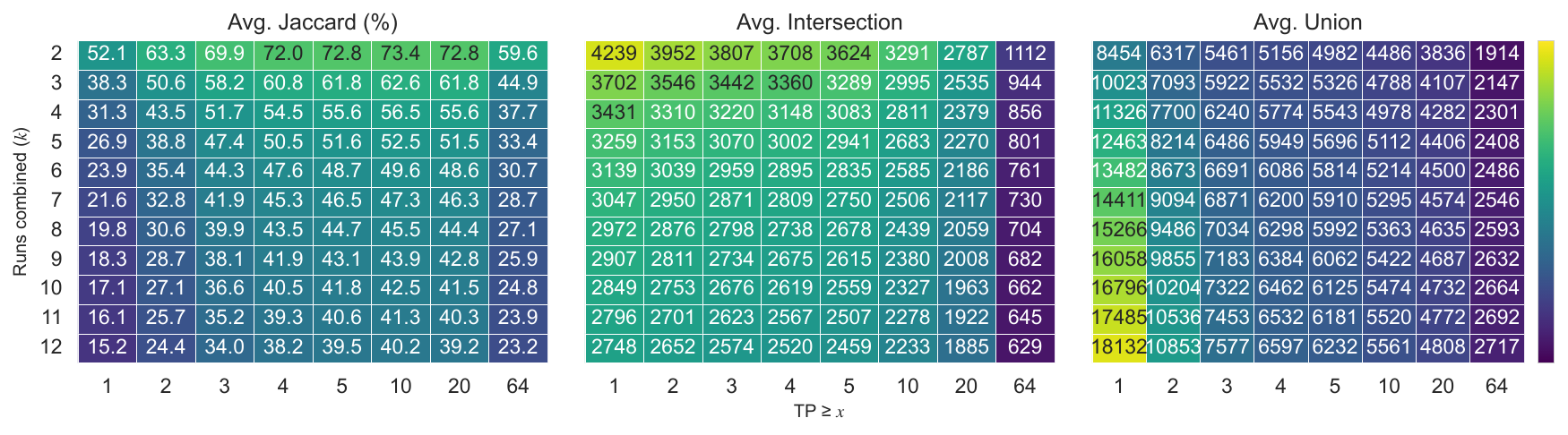}
    \caption{Reproducibility, stability, and coverage of \emph{all} LiRA positives across runs (identical settings) at a target FPR of 0.1\%.}
    \Description{Heatmaps showing reproducibility, stability, and coverage for all online LiRA detections at 0.1\% FPR across multiple identical-setting runs. 
                Agreement improves slightly compared to the 0.001\% setting, but coverage remains large and consistency still increases as more runs are combined.}
    \label{fig:tp_geq_all_identical_01}
\end{figure*}

At the more tolerant 0.1\% FPR (Fig.~\ref{fig:tp_geq_all_identical_01}), stability increased slightly, but the union remained high for higher Jaccard similarities and continued to increase as more runs were combined.

\begin{figure*}[t!]
    \centering
    \includegraphics[width=\linewidth]{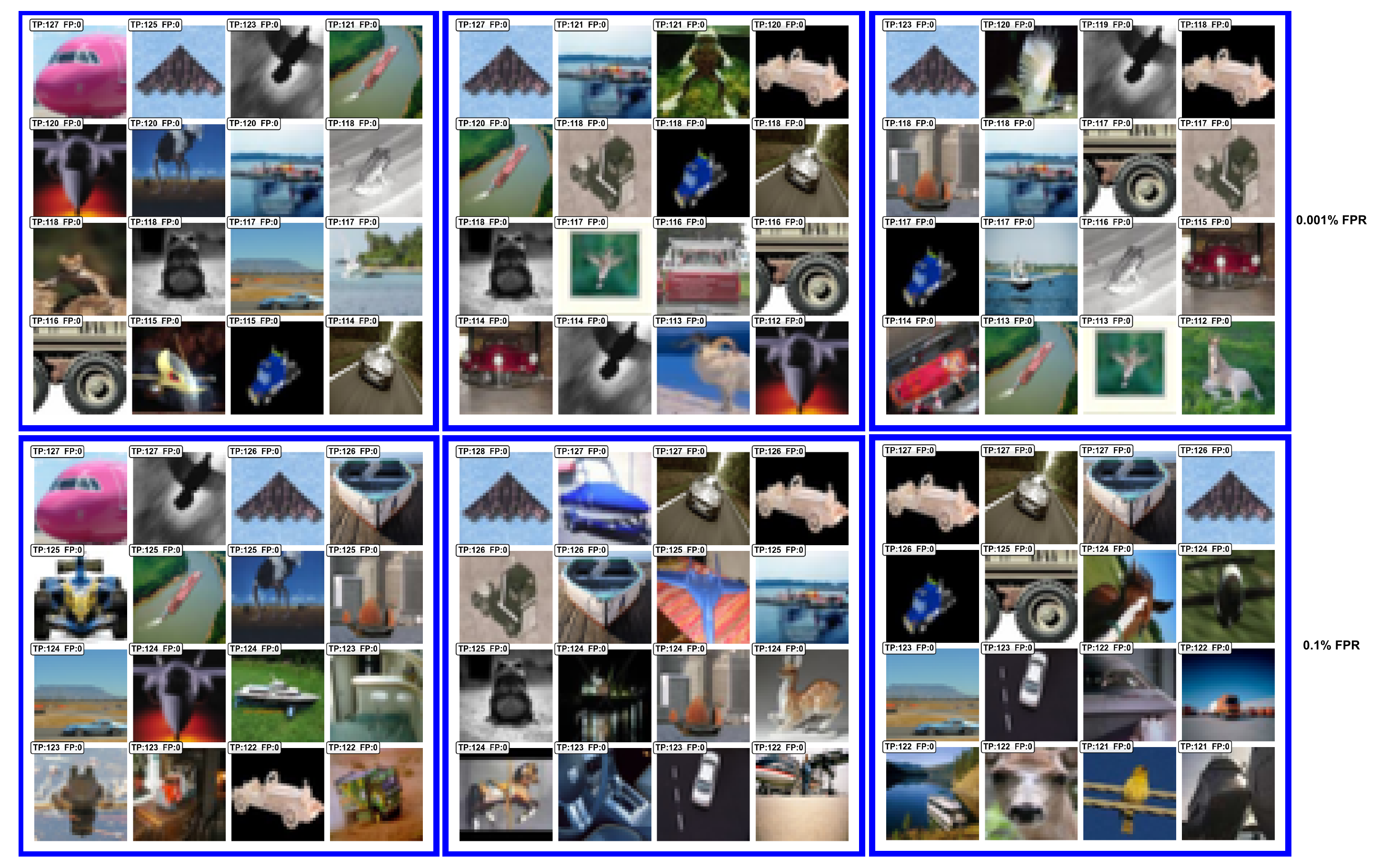}
    \caption{Top-16 most vulnerable samples across three seeds (columns) under two FPR settings (rows). 
    Substantial cross-FPR disagreement is visible: many samples appear in only one FPR setting, and those shared often change rank.}
    \Description{A grid comparing the top sixteen vulnerable samples identified across seeds under the two different FPR settings. 
                Many samples appear only at one FPR, and shared samples frequently change position, indicating substantial instability in which samples are deemed most vulnerable.}
    \label{fig:top_fpr_disagreement}
\end{figure*}

Fig.~\ref{fig:top_fpr_disagreement} compares the top-16 most vulnerable samples across three seeds at two FPR levels (0.001\% vs.\ 0.1\%).  
There is clear cross-FPR inconsistency: in at least half of the runs, the identity and ranking of the top samples differ between the two FPRs.  
This confirms that even ``top vulnerable samples'' are not stable across FPR choices and that vulnerability assessments depend strongly on random training factors and threshold calibration.

\begin{figure*}[t!]
    \centering
    \includegraphics[width=\linewidth]{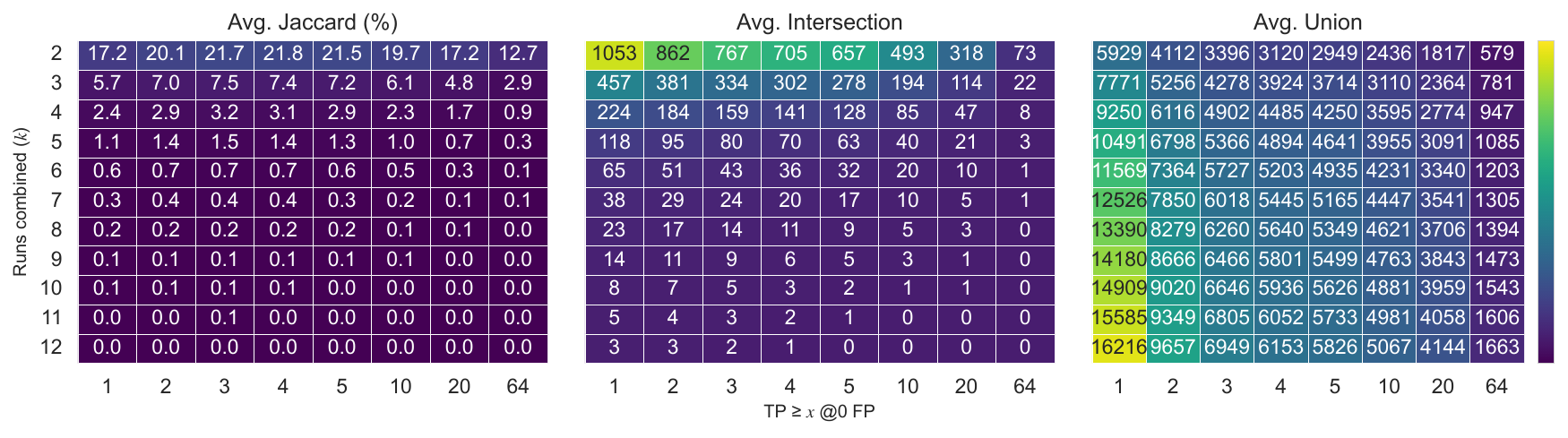}
    \caption{Reproducibility, stability, and coverage of zero-FP LiRA positives across runs (identical settings) for 0.1\%FPR.}
    \Description{The figure presents a grid of heatmaps that summarize reproducibility, stability, and coverage for zero-false-positive LiRA detections at 0.1\% FPR under identical training settings. 
        Each row corresponds to a different number of combined runs, and each column corresponds to a support threshold, defined as the minimum number of agreeing IN shadow models (out of 128) required for a detection within a run. 
        The heatmaps show how reproducibility decreases as more runs are added, and how increasing the support threshold reduces the number of shared detections while also shrinking total coverage. 
        The visualization highlights the trade-off between requiring stronger within-run agreement and maintaining consistent detections across multiple runs.}
    \label{fig:tp_geq_identical_01}
\end{figure*}

Fig.~\ref{fig:tp_geq_identical_01} further shows that at 0.1\% FPR, the online LiRA variant no longer produced zero-FP detections once the number of combined runs exceeded 10.

Table~\ref{tab:fpr_vs_topq_mean_gap_expanded} reports reproducibility metrics for the alternative \texttt{mean gap} vulnerability score.
The qualitative trends mirror those observed for the \texttt{median gap} in the main text, but stability in the extreme tail is consistently weaker.

At $q=1\%$, pairwise tail Spearman drops from 53.45\% (median gap) to 45.26\% (mean gap), and variance increases (std.\ 20.49 vs.\ 13.79).
The results are weaker, but qualitatively consistent with the \texttt{median gap}.
Under strict multi-run agreement ($k=12$), the globally stable core is also smaller.
For example, at $q=2.24\%$, the mean gap yields a global intersection of 338 samples (union 2466, Jaccard 13.71\%), compared to 368 samples (union 2280, Jaccard 16.14\%) under the median gap.

These differences indicate that the mean aggregation is more sensitive to variability across shadow models, leading to greater dispersion in the extreme tail.
In contrast, the median gap is more robust to outlier shadow likelihood ratios, resulting in a more stable ranking signal across runs.
Overall, while both scores preserve a population-level ordering, the median gap provides systematically stronger extreme-tail stability and tighter global agreement.
\begin{table*}[t!]
\caption{Reproducibility of ranking-based sets (Top-$q\%$ by \texttt{mean gap} and FPR-thresholded sets)}
\label{tab:fpr_vs_topq_mean_gap_expanded}
\centering
\small
\resizebox{\linewidth}{!}{
\begin{tabular}{llcccccccc}
\toprule
& & \multicolumn{4}{c}{Pairwise ($k=2$ runs)} & \multicolumn{4}{c}{All-runs ($k=12$ runs)} \\
\cmidrule(lr){3-6}\cmidrule(lr){7-10}
Setting & $T_q$ 
& Tail $\rho$ (\%) & $|\cap|$ & $|\cup|$ & Jaccard (\%)
& Tail $\rho$ (\%) & $|\cap|$ & $|\cup|$ & Jaccard (\%) \\
\midrule
0.10  & 50    & 39.41$_{\pm 26.57}$ & 23$_{\pm 9}$    & 77$_{\pm 9}$    & 31.57$_{\pm 16.19}$ & 55.45 & 4     & 171   & 2.34  \\
0.25  & 125   & 43.13$_{\pm 17.97}$ & 65$_{\pm 21}$   & 185$_{\pm 21}$  & 37.31$_{\pm 16.84}$ & 52.66 & 16    & 364   & 4.40  \\
0.50  & 250   & 45.16$_{\pm 19.14}$ & 137$_{\pm 36}$  & 363$_{\pm 36}$  & 39.20$_{\pm 15.00}$ & 47.11 & 42    & 708   & 5.93  \\
0.75  & 375   & 44.03$_{\pm 20.75}$ & 222$_{\pm 48}$  & 528$_{\pm 48}$  & 43.17$_{\pm 13.82}$ & 46.10 & 74    & 975   & 7.59  \\
1.00  & 500   & 45.26$_{\pm 20.49}$ & 305$_{\pm 61}$  & 695$_{\pm 61}$  & 45.09$_{\pm 13.49}$ & 46.37 & 111   & 1255  & 8.84  \\
2.00  & 1000  & 50.68$_{\pm 17.86}$ & 677$_{\pm 103}$ & 1323$_{\pm 103}$& 52.12$_{\pm 12.38}$ & 50.35 & 296   & 2228  & 13.29 \\
2.24  & 1121  
        & 52.01$_{\pm 17.50}$ & 768$_{\pm 110}$ & 1474$_{\pm 110}$ 
        & 53.03$_{\pm 11.86}$ & 50.63 & 338 & 2466 & 13.71 \\
3.00  & 1500  & 54.92$_{\pm 16.88}$ & 1056$_{\pm 137}$& 1944$_{\pm 137}$& 55.06$_{\pm 11.17}$ & 53.99 & 483   & 3153  & 15.32 \\
4.00  & 2000  & 57.92$_{\pm 15.66}$ & 1464$_{\pm 170}$& 2536$_{\pm 170}$& 58.42$_{\pm 10.81}$ & 57.64 & 717   & 3931  & 18.24 \\
5.00  & 2500  & 59.80$_{\pm 14.97}$ & 1882$_{\pm 201}$& 3118$_{\pm 201}$& 61.02$_{\pm 10.47}$ & 58.42 & 987   & 4708  & 20.96 \\
10.00 & 5000  & 69.03$_{\pm 12.33}$ & 4048$_{\pm 332}$& 5952$_{\pm 332}$& 68.51$_{\pm 9.19}$  & 67.57 & 2459  & 8300  & 29.63 \\
10.38 & 5190  
        & 69.59$_{\pm 12.17}$ & 4211$_{\pm 345}$ & 6169$_{\pm 345}$ 
        & 68.76$_{\pm 9.21}$ & 68.04 & 2551 & 8572 & 29.76 \\
20.00 & 10000 & 77.69$_{\pm 10.46}$ & 8624$_{\pm 532}$& 11376$_{\pm 532}$& 76.18$_{\pm 7.99}$ & 76.75 & 5902  & 14671 & 40.23 \\
30.00 & 15000 & 82.75$_{\pm 9.38}$  & 13141$_{\pm 739}$&16859$_{\pm 739}$& 78.28$_{\pm 7.59}$ & 81.62 & 9409  & 21296 & 44.18 \\
40.00 & 20000 & 85.86$_{\pm 7.89}$  & 17537$_{\pm 1076}$&22463$_{\pm 1076}$& 78.45$_{\pm 8.11}$& 85.15 & 12588 & 28556 & 44.08 \\
50.00 & 25000 & 87.70$_{\pm 6.61}$  & 21537$_{\pm 1431}$&28463$_{\pm 1431}$& 76.07$_{\pm 8.25}$& 87.29 & 15211 & 38180 & 39.84 \\
60.00 & 30000 & 88.95$_{\pm 5.47}$  & 25074$_{\pm 1544}$&34926$_{\pm 1544}$& 72.10$_{\pm 7.15}$& 88.58 & 17142 & 46914 & 36.54 \\
70.00 & 35000 & 89.67$_{\pm 4.59}$  & 28987$_{\pm 1397}$&41013$_{\pm 1397}$& 70.86$_{\pm 5.56}$& 89.31 & 18717 & 49776 & 37.60 \\
80.00 & 40000 & 89.10$_{\pm 4.47}$  & 34151$_{\pm 943}$ &45849$_{\pm 943}$ & 74.56$_{\pm 3.51}$& 89.92 & 20882 & 49997 & 41.77 \\
90.00 & 45000 & 86.28$_{\pm 6.18}$  & 41099$_{\pm 386}$ &48901$_{\pm 386}$ & 84.06$_{\pm 1.46}$& 90.15 & 25911 & 50000 & 51.82 \\
100.00& 50000 & 80.46$_{\pm 10.45}$ & 50000$_{\pm 0}$   &50000$_{\pm 0}$   &100.00$_{\pm 0.00}$& 80.46 & 50000 & 50000 &100.00 \\
\midrule
\multicolumn{10}{l}{\textbf{FPR-thresholded sets}}\\
FPR $=0.001\%$ & --- & --- & 954  & 2142  & 49.80 & --- & 367  & 4833  & 7.60  \\
FPR $=0.1\%$   & --- & --- & 4239 & 8454  & 52.10 & --- & 2748 & 18132 & 15.20 \\
\bottomrule
\end{tabular}}
\end{table*}

\begin{figure}[t!]
    \centering
    \includegraphics[width=\linewidth]{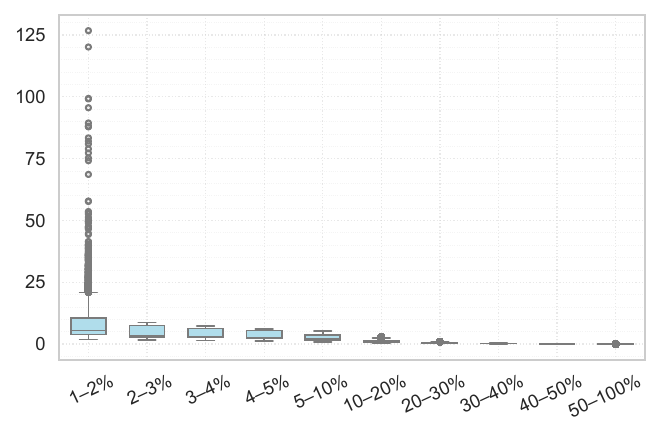}
    \caption{Distribution of \texttt{median gap} vulnerability scores within rank-percentile bins (bins computed per run; values pooled over 12 independent runs).}
    \Description{
    Boxplots of per-sample median-gap vulnerability scores grouped into rank-percentile bins (computed per run and pooled across 12 runs). 
    The top 1–2\% bins exhibit substantially larger dispersion and right-skew compared to lower-percentile bins, 
    illustrating that extreme-tail vulnerability scores are more sensitive to run-to-run variability.
    }
    \label{fig:median_gap_bins}
\end{figure}

To understand the source of instability at the score level, 
Figure~\ref{fig:median_gap_bins} plots the \texttt{median gap} score distribution within rank-percentile bins. 
Score dispersion was highest in the top 1--2\% bins and decreased rapidly beyond 5--10\%. 
In this sparse and high-variance regime, small training perturbations can reshuffle the ordering, whereas denser regions exhibit more stable rankings.

\begin{figure*}[t!]
    \centering
    \includegraphics[width=\linewidth]{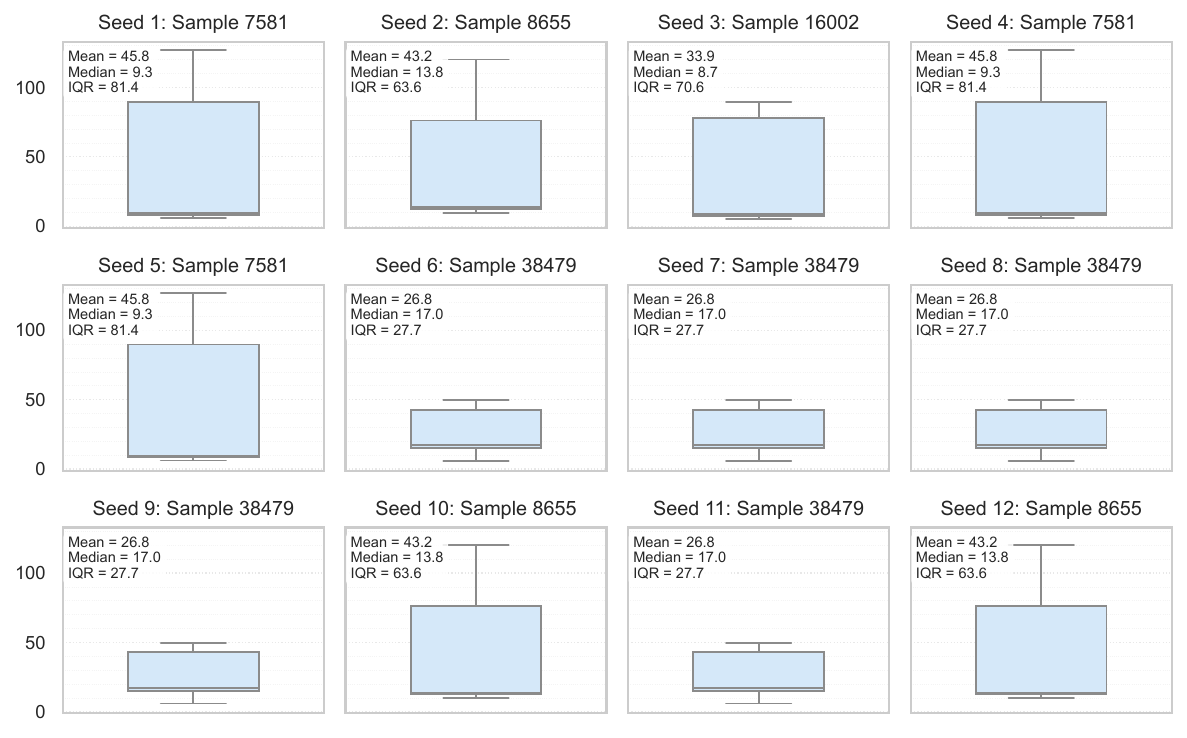}
    \caption{For each run, the top-1 most vulnerable sample is selected,
        and its \texttt{median gap} score is evaluated across all 12 retrainings.
        Large interquartile ranges and strong right-skew indicate substantial cross-run
        variance in extreme vulnerability magnitude.
        While the top-1 identity rotates among a small set of recurring candidates,
        their scores vary markedly across runs, confirming sensitivity at the extreme boundary.}
    \Description{
        Boxplots showing the distribution of median-gap scores for the sample that was ranked top-1 in each run, evaluated across all 12 retrainings. 
        Each boxplot corresponds to one run’s top-1 sample. 
        Large interquartile ranges and right-skew indicate that extreme vulnerability magnitude is not consistently preserved across runs.
        }
    \label{fig:top1_across_runs}
\end{figure*}

Fig.~\ref{fig:top1_across_runs} selects, for each run, the top-1 most vulnerable sample (\texttt{median gap}) and evaluates its score across all 12 runs.
We observe substantial cross-run dispersion, with large interquartile ranges and strong right-skew; in several cases, the median is far smaller than the mean, indicating that extreme values occur only in a subset of runs.
Thus, even the most vulnerable record in a given run does not consistently exhibit extreme vulnerability magnitude under retraining variability.

\begin{figure}[t!]
    \centering
    \includegraphics[width=\linewidth]{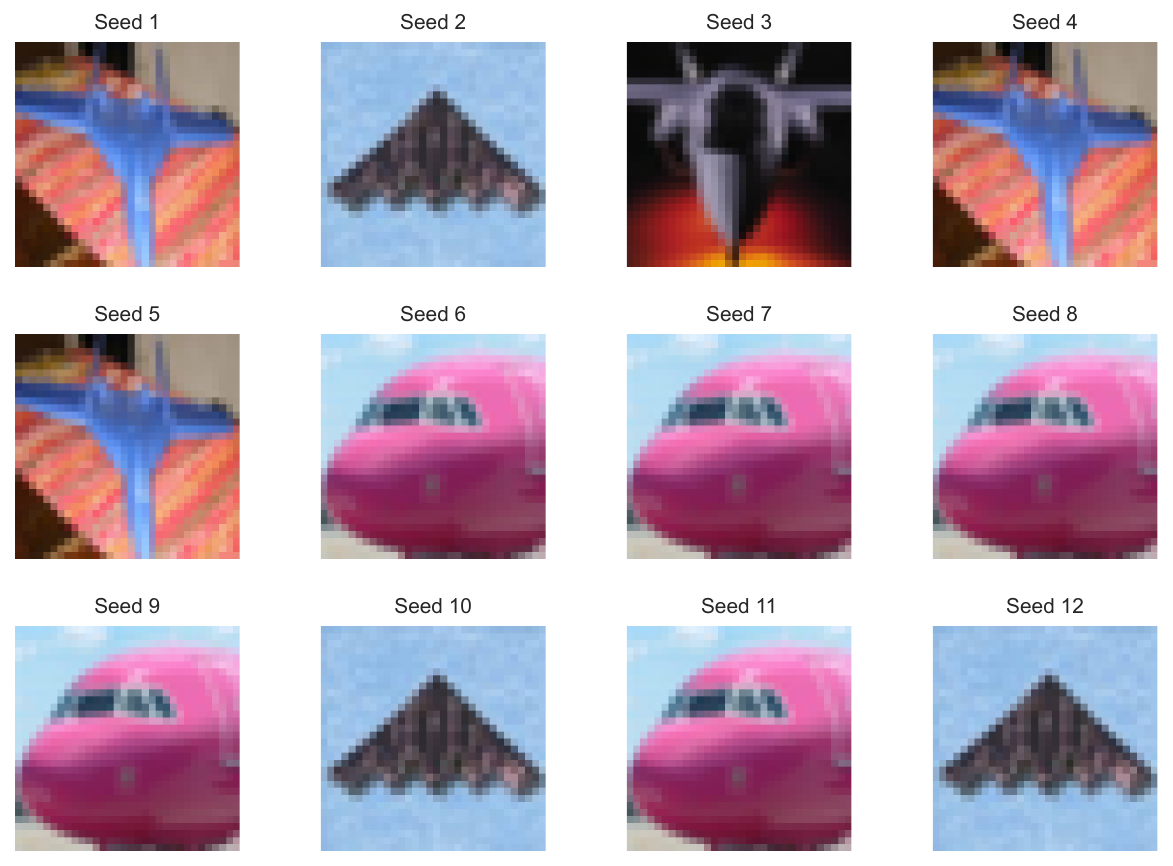}
    \caption{\textbf{Top-1 identity across runs.} For each seed (panel), we display the single most vulnerable sample in that run (highest \texttt{median gap}).        Repeated appearances of the same image across seeds indicate overlap in the extreme tail.}
    \Description{
        A 3×4 grid showing the single most vulnerable sample (highest median-gap score) identified in each of 12 independent runs. 
        Each panel corresponds to one run. 
        Repeated images across panels indicate overlap in top-1 identity, while differing images illustrate instability in extreme-tail selection.
        }
    \label{fig:top1_samples_grid}
\end{figure}

Fig.~\ref{fig:top1_samples_grid} shows the top-1 sample from each of the 12 runs
(\texttt{median gap}), arranged in a $3\times 4$ grid.

\section{Ablation Study on CIFAR-10}
\label{app:ablation}

We conducted a comprehensive ablation to isolate the effects of augmentation, dropout, and weight decay on the privacy–utility trade-off. 
All experiments used WideResNet-28-2 in CIFAR-10 and were evaluated with online LiRA (fixed variance) at 0.1\% FPR using target-based thresholds. 
Table~\ref{tab:ablation_full} reports complete results across 22 configurations.

\paragraph{Augmentation effect} 
Minimal augmentation (FLP+CRP) led to severe overfitting (loss ratio 713.8) and high vulnerability (TPR 8.47\%). 
Adding Cutout substantially mitigated both effects (loss ratio 26.1, TPR 5.93\%). 
Introducing stronger augmentations (rotation, ColorJitter, and CutMix) further improved generalization and privacy (loss ratio 1.35, TPR 0.78\%), achieving more than a sevenfold reduction in vulnerability compared to Cutout alone. 
Across configurations, CutMix consistently outperformed MixUp in privacy protection (TPR 1.87\% vs.\ 1.93\%), because its spatially coherent patches discourage pixel-level memorization.
Therefore, we selected CutMix for our final configurations.

\paragraph{Dropout effect}
Increasing dropout monotonically decreased TPR, 
though diminishing returns appear beyond 25\%. 
Raising dropout from 25\% to 50\% reduced TPR by only 1.2\% but lowered accuracy by 2.2\%. 
The 25\% configuration thus offers an effective balance between privacy (2.24\% TPR) and utility (92.2\% accuracy).

\paragraph{Weight decay effect}
Weight decay exhibited a narrow effective range ($5{\times}10^{-4}$ to $10^{-2}$). 
Lower values ($5{\times}10^{-5}$) provided insufficient regularization (loss ratio 17.6), 
while excessive decay ($5{\times}10^{-2}$) collapsed the model accuracy to 18.9\%. 
A value of $10^{-3}$ offered near-optimal stability and privacy.

\paragraph{Combined effects}
The best configuration (FLP+CRP+ROT+JTR +CMX, 10\% dropout, $10^{-3}$ weight decay) achieved 1.21\% TPR with 93.2\% test accuracy, approximately a 5$\times$ reduction in attack success compared to the baseline (CRP+FLP+CUT, 0\% dropout) while maintaining accuracy. 
This demonstrates that privacy and utility are not inherently conflicting when anti-overfitting measures are properly combined.

\begin{table*}[t!]
\centering
\caption{Complete ablation study: WideResNet-28-2 on CIFAR-10 with online LiRA (fixed variance) at 0.1\% FPR, target-based threshold. Configurations sorted by TPR (descending) within each technique category.}
\label{tab:ablation_full}
\scriptsize
\resizebox{\linewidth}{!}{
\begin{tabular}{lccccccc} 
\toprule
Augmentation                & DRP (\%) & WD   & Train Loss & Test Loss & Loss Ratio & Test Acc (\%) & TPR (\%)  \\ 
\midrule
\multicolumn{8}{l}{\textit{Augmentation variations}}                                                            \\
CRP+FLP                     & 0        & 5e-4 & 0.0004     & 0.2855    & 713.75     & 93.03         & 8.47      \\
CRP+FLP+ROT                 & 0        & 5e-4 & 0.0010     & 0.2389    & 238.90     & 93.38         & 8.01      \\
CRP+FLP+CUT (LiRA)          & 0        & 5e-4 & 0.0079     & 0.2061    & 26.09      & 93.76         & 5.93      \\
CRP+FLP+ROT+CUT             & 0        & 5e-4 & 0.0241     & 0.2206    & 9.15       & 93.08         & 4.20      \\
CRP+FLP+ROT+CUT+JTR+MIX     & 0        & 5e-4 & 0.1221     & 0.2421    & 1.98       & 92.66         & 1.93      \\
CRP+FLP+ROT+JTR+CMX         & 0        & 5e-4 & 0.1804     & 0.2864    & 1.59       & 93.69         & 1.87      \\
CRP+FLP+ROT+CUT+JTR+MIX+CMX & 0        & 5e-4 & 0.1700     & 0.2695    & 1.59       & 92.14         & 1.19      \\
CRP+FLP+ROT+CUT+JTR+CMX     & 0        & 5e-4 & 0.2381     & 0.3225    & 1.35       & 91.31         & 0.78      \\ 
\cmidrule(lr){1-8}
\multicolumn{8}{l}{\textit{Dropout variations}}                                                                 \\
CRP+FLP+CUT                 & 10       & 5e-4 & 0.0440     & 0.2228    & 5.06       & 92.83         & 3.21      \\
CRP+FLP+CUT                 & 25       & 5e-4 & 0.0777     & 0.2387    & 3.07       & 92.16         & 2.24      \\
CRP+FLP+CUT                 & 35       & 5e-4 & 0.1087     & 0.2567    & 2.36       & 91.50         & 1.67      \\
CRP+FLP+CUT                 & 50       & 5e-4 & 0.1719     & 0.3009    & 1.75       & 89.97         & 1.01      \\ 
\cmidrule(lr){1-8}
\multicolumn{8}{l}{\textit{Weight decay variations}}                                                            \\
CRP+FLP+CUT                 & 0        & 5e-5 & 0.0146     & 0.2565    & 17.57      & 92.56         & 4.77      \\
CRP+FLP+CUT                 & 0        & 5e-3 & 0.1109     & 0.2329    & 2.10       & 92.21         & 1.39      \\
CRP+FLP+CUT                 & 0        & 1e-2 & 0.2368     & 0.3276    & 1.38       & 89.25         & 0.54      \\
CRP+FLP+CUT                 & 0        & 5e-2 & 2.1562     & 2.1567    & 1.00       & 18.92         & 0.10      \\ 
\cmidrule(lr){1-8}
\multicolumn{8}{l}{\textit{Combined strategies (sorted by TPR descending)}}                                     \\
CRP+FLP+CMX                 & 25       & 5e-4 & 0.1828     & 0.3023    & 1.65       & 92.25         & 1.61      \\
CRP+FLP+ROT+JTR+CMX         & 10       & 5e-4 & 0.2142     & 0.3094    & 1.44       & 93.00         & 1.37      \\
CRP+FLP+ROT+JTR+CMX         & 10       & 1e-3 & 0.2131     & 0.3029    & 1.42       & 93.18         & 1.21      \\
CRP+FLP+CMX                 & 50       & 5e-4 & 0.2586     & 0.3615    & 1.40       & 89.95         & 0.93      \\
CRP+FLP+ROT+CUT+JTR+CMX     & 25       & 5e-4 & 0.3169     & 0.3862    & 1.22       & 89.03         & 0.52      \\
CRP+FLP+ROT+CUT+JTR         & 25       & 5e-3 & 0.3319     & 0.3915    & 1.18       & 86.65         & 0.33      \\
\bottomrule
\end{tabular}
}
\end{table*}

\paragraph{Transfer learning and architecture choice}
\label{app:tl_choice}
To isolate the impact of transfer learning, 
we first compared ResNet-18 trained from scratch and with ImageNet pretraining on CIFAR-10. 
As shown in Table~\ref{tab:cifar10-tl_effect}, TL reduced LiRA’s TPR by more than an order of magnitude while improving loss ratio and accuracy. 
We also observed that pretrained EfficientNet-V2 yields a comparable drop in attack success, but achieves higher test accuracy. 
Consequently, we employed EfficientNet-V2 for the main experiments, as a pragmatic practitioner will prefer higher model utility.
\begin{table*}[t!]
\centering
\caption{CIFAR-10 results under the online LiRA attack. 
The table isolates the effect of TL.
Pretrained EfficientNet-V2 (EN-V2) yields a similar effect on attack as pretrained ResNet-18 (RN-18) while providing higher accuracy.}
\label{tab:cifar10-tl_effect}
\small
\begin{tabular}{lcccc}
\toprule
Benchmark & TPR@0.001\% FPR (\%) & TPR@0.1\% FPR (\%) & Loss Ratio & Test Acc (\%) \\ 
\midrule
\multirow{1}{*}{Baseline (RN-18)} 
& 3.956$_{\pm 1.061}$ & 10.268$_{\pm 0.555}$ & 71.0 & 93.63 \\
\cmidrule(lr){1-5}
\multirow{1}{*}{AOF (RN-18)} 
& 0.248$_{\pm 0.198}$ ($\times$16) & 2.723$_{\pm 0.683}$ ($\times$3.8) & 1.88 & 94.09 \\
\cmidrule(lr){1-5}
\multirow{1}{*}{AOF+TL (RN-18)} 
& 0.076$_{\pm 0.055}$ ($\times$52) & 0.782$_{\pm 0.171}$ ($\times$13) & 1.24 & 95.10 \\
\cmidrule(lr){1-5}
\multirow{1}{*}{AOF+TL (EN-V2)} 
& 0.065$_{\pm 0.061}$ ($\times$61) & 0.521$_{\pm 0.128}$ ($\times$20) & 1.36 & 97.00 \\
\bottomrule
\end{tabular}
\end{table*}

\section{Additional Training Details}
\label{app:additional_setting}

Baseline vision models followed LiRA~\cite{carlini2022membership} with data augmentations consisting of random horizontal flip, reflection-padded 4px crop, and Cutout~\cite{devries2017improved}; no dropout; and weight decay $5{\times}10^{-4}$.  
For AOF benchmarks, we explored multiple augmentation strategies:
rotation ($\pm 15^\circ$ for CIFAR-10/100; $\pm 10^\circ$ for GTSRB) applied with 50\% probability; ColorJitter with brightness, contrast, and saturation jitter of $\pm0.4$, and hue jitter of $\pm0.1$ (parameters reduced by half for GTSRB due to its more constrained color distribution); 
Cutout~\cite{devries2017improved} (16$\times$16 patches) for baseline configurations; 
MixUp~\cite{zhang2018mixup} that linearly interpolates pairs of training samples and their labels with mixing coefficient $\lambda \sim \mathrm{Beta}(1.0, 1.0)$; CutMix~\cite{yun2019cutmix} that replaces rectangular regions between image pairs while mixing labels proportionally to area, applied with 50\% probability and $\lambda \sim \mathrm{Beta}(1.0, 1.0)$. 
The best AOF combination for vision tasks consists of horizontal flip, reflection-padded crop, rotation, ColorJitter, and CutMix, with 10\% dropout and weight decay of $10^{-3}$ (see Appendix~\ref{app:ablation}). 
For TL, we used the same augmentations but increased dropout to 25\% and weight decay to $5{\times}10^{-2}$ to account for the capacity of the pretrained model. 
For Purchase-100, we compared no dropout with $5{\times}10^{-4}$ weight decay (baseline) against 50\% dropout with $10^{-3}$ weight decay (AOF). 
No augmentations were applied on tabular data.

Experiments were conducted on Windows~11 Home with an Intel\textsuperscript{\textregistered} Core\texttrademark~i7-12700 (12 cores), 32GB RAM, and an NVIDIA GeForce RTX~4080 (16GB VRAM).

\end{document}